\documentclass[preprint,12pt]{elsarticle}

\usepackage{amssymb}
\usepackage{amsmath}
\usepackage{url}
\usepackage{hyperref}
\usepackage{multirow}
\usepackage{subfigure}
\usepackage{threeparttable}
\usepackage{booktabs} 
\usepackage{makecell}
\usepackage{color}
\usepackage[T1]{fontenc}

\journal{Journal of High Energy Astrophysics}

\begin{document}

\begin{frontmatter}



\title{{\it Fermi}-LAT and FAST observation of the gamma-ray binary HESS J0632+057}


\author[1,2]{Yanlv Yang}
\affiliation[1]{Department of Astronomy, University of Science and Technology of China, Hefei 230026, China}
\affiliation[2]{School of Astronomy and Space Science, University of Science and Technology of China, Hefei 230026, China}
\author[3]{Dengke Zhou}
\affiliation[3]{Research Center for Astronomical Computing, Zhejiang Laboratory, Hangzhou 31112l, China,}
\author[1,2]{Zihao Zhao}

\author[1,2]{Jian Li\corref{cor1}}
\cortext[cor1]{Corresponding author}
\ead{jianli@ustc.edu.cn}
\author[7,8,9]{Diego F. Torres}
\affiliation[7]{Institute of Space Sciences (ICE, CSIC), Campus UAB, E-08193 Barcelona, Spain}
\affiliation[8]{Institut d’Estudis Espacials de Catalunya (IEEC), E-08034 Barcelona, Spain}
\affiliation[9]{Institució Catalana de Recerca i Estudis Avanc˛ats (ICREA), E-08101 Barcelona, Spain}
\author[10,11]{Pei Wang}
\affiliation[10]{National Astronomical Observatories, Chinese Academy of Sciences, Beijing 100101, China}
\affiliation[11]{Institute for Frontiers in Astronomy and Astrophysics, Beijing Normal University, Beijing 102206, China}

\begin{abstract}
Using 15 years of data from the {\it Fermi} Large Area Telescope ({\it Fermi}-LAT), we performed a comprehensive analysis on the gamma-ray binary HESS J0632+057. 
Its spectrum in 0.1–300 GeV band is well described by a power law model with an index of $2.40\pm0.16$, leading to an energy flux of (5.5$\pm$1.6$)\times$ 10$^{-12}$ erg cm$^{-2}$ s$^{-1}$.
The GeV Spectral Energy Distribution (SED) of HESS J0632+057 hints for a spectral turn-over between $\sim$10-100 GeV.
Orbital analysis reveals a flux enhancement during the phase range of 0.2-0.4, consistent with the X-ray and TeV light curves, indicating an origin of a common particle population. 
We carried out six deep radio observations on HESS J0632+057 with the Five-hundred-meter Aperture Spherical Telescope (FAST), evenly distributed across its orbit, reaching a detection sensitivity of 2$\mu$Jy.
However, no radio pulsation was detected within these observations. 
The absence of radio pulsation may be attributed to the dense stellar wind environment of HESS J0632+057.
\end{abstract}




\end{frontmatter}



\section{Introduction} \label{sec:intro}
Gamma-ray binaries are binary systems emitting most of their electromagnetic radiation in the MeV to TeV band, typically consisting of a compact object (e.g. neutron star or black hole) and a massive star (Be or O type)\citep[see][]{2013A&ARv..21...64D}. 
There are 7 known in our galaxy: LS 5039 \citep{2005Sci...309..746A, 2006A&A...460..743A}, LS I+61 303 \citep{2009ApJ...701L.123A}, PSR B1259$-$63\citep{2005A&A...442....1A}, HESS J0632+057 (see below), 1FGL J1018.6-5856 \citep{2012Sci...335..189F}, 4FGL J1405.1-6119\cite{2019ApJ...884...93C}, PSR J2032+4127 \citep{2018ApJ...867L..19A} and one in the Large Magellanic Cloud: LMC P3 \citep{2016ApJ...829..105C}. 
Though Cyg X-1 \citep{2007ApJ...665L..51A} and Cyg X-3 \citep{2009Sci...326.1512F} have also been detected in gamma-rays, their emissions peak in the X-ray band and thus not counted as gamma-ray binary.

The gamma-ray binary HESS J0632+057 was identified as a TeV point source by H.E.S.S during the observation of the Monoceros Loop supernova remnant (SNR) in 2004$-$2005 \citep{2007A&A...469L...1A}. 
It is also the only gamma-ray binary observable both in the northern and southern hemisphere. 
Followed by a 26 ks {\it XMM-Newton} observation, an X-ray source XMMU J063259.3+05480 {is found to be positionally coincident with} HESS J0632+057, showing a gradual decline in X-ray flux \citep{2009ApJ...690L.101H}. 
The low coincidence probability ($\sim 10^{-4}$, \cite{2007A&A...469L...1A}) indicates it to be the X-ray counterpart of HESS J0632+057.
A B0Vpe star star MWC 148 (HD 259440) was reported as its optical counterpart \citep{2009ApJ...690L.101H}. 
The mass and radius of MWC 148 were estimated to be in the range of $13.2-19.0 M_\odot$ and $6.0-9.6 R_\odot$, respectively, and it is located $1.1-1.7$ kpc away \citep{2010ApJ...724..306A}.  
The absence of TeV emission {as reported in VERITAS observations} indicated there might be a flux variation of HESS J0632+057 \citep{2009ApJ...698L..94A}, supporting its nature as a gamma-ray binary. 
The {\it ROSAT} source 1RXS J063258.3+054857 \citep{2000IAUC.7432....3V}, the EGRET source 3EG J0634+0521 \citep{2007A&A...469L...1A} and {\it Fermi}-LAT source 4FGL 0632.8+0550 \citep{2017ApJ...846..169L,2020ApJS..247...33A} have also been identified as the X-ray and GeV counterpart of HESS J0632+057.

{Gamma-ray binaries typically exhibit multi-wavelength periodic variability, arising from the interaction between the compact object and the massive companion star or their outflows, which is linked to the orbital motion of the system.}
The orbital period of HESS J0632+057 has been estimated via multi-wavelength observations. 
\emph{Swift/XRT} observations showed variations in different timescales, from several days to a month \citep{2010ApJ...708L..52F}, suggesting an orbital period to be either 36-40 days or above 54 days. 
%
%
Years long {\it Swift}/XRT monitoring by \cite{2011ApJ...737L..11B} gave a measurement on orbital period of $321\pm5$ d, establishing the definitive confirmation of its binary nature. 
The long-term X-ray dataset was used to derive
a more precise period of $315 ^{+6}_{-4}$ days by \cite{2014ApJ...780..168A}. It was further refined to 317.3$\pm$0.3 days by \cite{2021ApJ...923..241A} with 15 years of X-ray data including {\it Swift}-XRT, {\it Chandra}, {\it XMM-Newton}, {\it NuSTAR}, and {\it Suzaku}.
The orbital solution from optical spectroscopic analyses supports an eccentricity $e=0.83\pm0.08 $ and a mass of
the compact object in the range 1.3$-$7.1  $M_\odot$ \citep{2012MNRAS.421.1103C}.
Another orbital configuration was proposed  by \cite{2018PASJ...70...61M}, with a slightly lower eccentricity of $\sim$0.6 and a mass of the compact object less than 2.5 $M_\odot$. 
Combining observations from the Southern African Large Telescope (SALT) and those in the previous two papers, \cite{2025MNRAS.536..166M} 
reported two eccentricities of $e=0.40\pm0.08$ and $e=0.75\pm0.24$, respectively.

Phase-folded light curves of HESS J0632+057 in both X-rays and gamma-rays, as presented by \cite{2021ApJ...923..241A}, show synchronized orbital variability patterns, with a distinct peak occurring between orbital phases 0.2 and 0.4.
Meanwhile, a {\it Fermi}-LAT analysis of broad orbital phase by \cite{2017ApJ...846..169L} revealed a significantly higher flux in phase 0-0.5 than the other half (0.5-1) in GeV observations, consistent with X-ray and TeV results. 
{The orbital phase was only split in two intervals (0-0.5 and 0.5-1) due to the limited statistics available at the time.}

{The origins of high energy emission from HESS J0632+057, as well as gamma-ray binaries in general, are still under debate.
The gamma-ray emissions are proposed to originate from the interaction between the relativistic outflow of the compact object and either the non-relativistic wind or the radiation field of the massive companion star \citep{2020mbhe.confE..45C}. This process may involve either accretion onto the compact object or the dissipation of its rotational energy. 
PSR B1259-63 \citep{2015MNRAS.454.1358C,2019A&A...627A..87C} and PSR J2032+4127 \citep{2017ApJ...836..241T, 2022A&A...658A.153C} have been characterized to have a young rotation powered pulsar producing pulsar wind which interacts with the stellar wind of the companion. 
The existence of a pulsar would reveal itself with a possible pulse signal, thus we can verify the mechanism of gamma-ray emission behind the binary system. }
However, no pulsation has been identified in HESS J0632+057 in the past.

In this paper, we present an updated analysis of HESS J0632+057 in the GeV band using 15 years of observations from {\it Fermi}-LAT. 
To explore the nature of its compact object, we report on a search for periodic radio pulse signals on HESS J0632+057 conducted through six deep FAST observations. 
In \autoref{observations}, we describe the {\it Fermi}-LAT data used and the related likelihood analysis method. 
A nearby bright gamma-ray pulsar is gated off as described in \autoref{gateoff}, followed by the presentation of gamma-ray emission results in \autoref{spectral}. 
The periodic pulse search conducted with FAST is described in \autoref{pulse}, and the conclusions are discussed in \autoref{conclusion}.
%
%

\section{{\it Fermi}-LAT Observations}
\label{observations}
We utilized data from Large Area Telescope (LAT) onboard the {\it Fermi Gamma-ray Space Telescope} to perform a gamma-ray analysis of HESS J0632+057.
Our analysis considered 15 years of LAT observations, spanning from August 4, 2008 to August 30, 2023, in accordance with the ephemeris of PSR J0633+0632 used in the later stages of our study.
The analysis was performed using the {\it Fermitools}\footnote{\url{https://fermi.gsfc.nasa.gov/ssc/data/analysis/software/}} with the P8R3 V2 instrument response functions (IRFs)\footnote{\url{https://fermi.gsfc.nasa.gov/ssc/data/analysis/documentation/Cicerone/Cicerone_LAT_IRFs/IRF_overview.html}}.
We selected photons from the "P8 Source" event class (evclass=128) and the "FRONT+BACK" event type (evtype=3). Photons were extracted within a 15-degree circular region of interest (ROI) in the energy range of 0.1–300 GeV. 
Additionally, a zenith angle cut of $<90^\circ$ was applied to minimize contamination from the Earth's limb.

We build the background source model from the 4FGL DR4 catalog\footnote{Data available in \url{https://fermi.gsfc.nasa.gov/ssc/data/access/lat/}} \citep{2023arXiv230712546B}, including gamma-ray sources within 20 degrees of HESS J0632+057 and fixing all spectral parameters except for sources within 5 degrees of HESS J0632+057. 
For these nearby sources, the prefactor and index parameters were allowed to vary to optimize the likelihood fit.
In addition, the Galactic diffuse emission model \emph{gll\_iem\_v07.fits} and isotropic diffuse model \emph{iso\_P8R3\_SOURCE\_V3\_v1} are included in the background model as well.
The gamma-ray spectral and spatial analyses were conducted using a binned maximum likelihood method \citep{1996ApJ...461..396M} implemented with the \emph{gtlike} tool.
The Test Statistics (TS) value is utilized to test the goodness of fit.
\begin{equation}
    \mathcal{TS}=-2ln\left(\frac{L_{max,0}}{L_{max,1}}\right)
\end{equation}

where $L_{max,0}$ is the maximum likelihood value when the target source is removed from the model ("null hypothesis") and $L_{max,1}$ represents the maximum likelihood value of the full model.
In the likelihood algorithm, the TS values of individual bins are summed to compute the overall TS, enabling a maximum likelihood fit across the gamma-ray energy band. The significance level is approximately the square root of TS value.

\section{Gating off gamma-ray pulsar PSR J0633+632}
\label{gateoff}

\begin{figure}[h]
    \centering
    \includegraphics[scale=0.4]{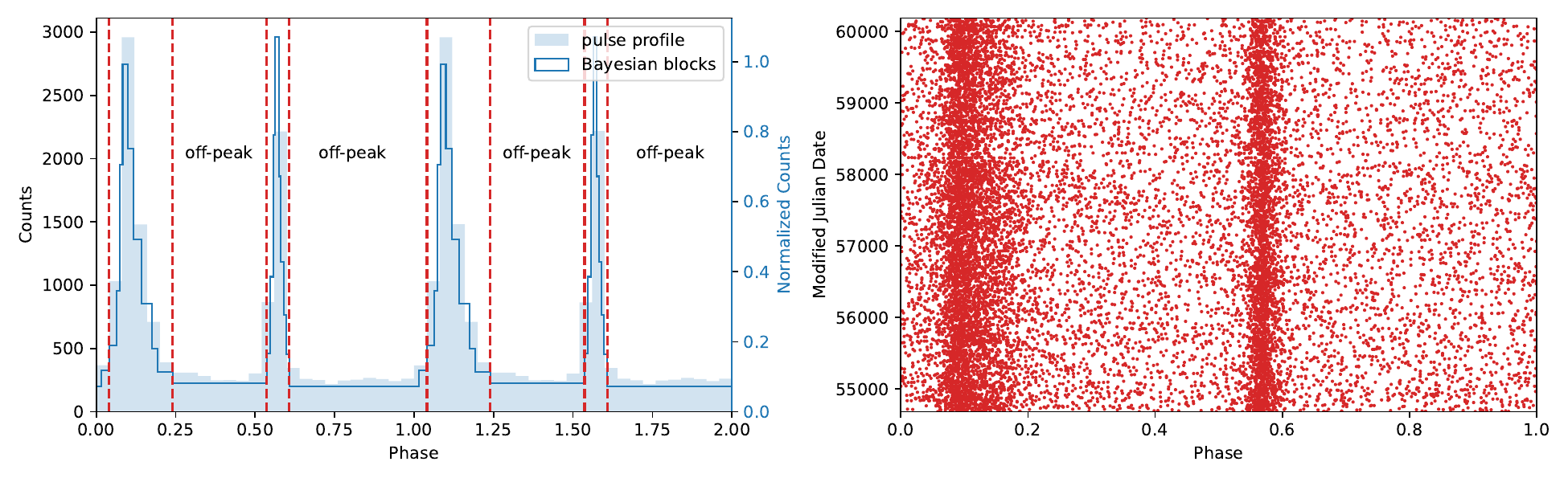}
    \caption{Timing analysis of PSR J0633+0632: \emph{left} panel: pulse profile shown in two periods. 
    The pulsating phases are selected with Bayesian block calculation, illustrated in red dashed lines, and off-peak phases $\phi$=0-0.0406, 0.24-0.5364 and 0.6075-1 are adopted for further analysis; \emph{right} panel: pulse phase for each event vs. time.}
    \label{psr_pulse}
\end{figure}

HESS J0632+057 is identified as 4FGL J0632.8+0550 in the LAT Fourth Source Catalog \citep[4FGL;][]{Abdollahi_2020}.
It is situated in a dense region.
A bright gamma-ray pulsar, PSR J0633+0632 (4FGL J0633.7+0632), is located just $\sim$0.76 degrees away from HESS J0632+057. 
The bright gamma-ray emission from PSR J0633+0632 may bring contamination to the analysis of HESS J0632+057. 
To mitigate this, we utilized its ephemeris to gate off the pulsating peaks through timing analysis. Photons within $0.6^\circ$ of PSR J0633+0632 in 100 MeV-300 GeV were selected.
Using an updated gamma-ray ephemeris of HESS J0632+057 contemporaneous with our {\it Fermi}-LAT observations (MJD 54682-MJD 60186), we calculated the pulse phase for each gamma-ray event over the observation period. 
The pulse profile and pulse phase for each event vs. time are shown in \autoref{psr_pulse}.

Using the pulse profile shown in \autoref{psr_pulse}, we determined the off-peak phases by performing a Bayesian block calculation.
For this study, we adopt the off-peak phase intervals of $\phi$ = 0–0.0406, 0.24–0.5364, and 0.6075–1. 
To account for these phase cuts in subsequent off-peak analyses, the prefactor parameters were scaled to the relative width of the phase
interval 0.7295.

\section{Gamma-ray emission of HESS J0632+057}
\label{spectral}

We carried out gamma-ray data analysis of HESS J0632+057 during the off-peak of PSR J0633+0632. 
The gamma-ray emission from HESS J0632+057 was spatially localized using \emph{Fermipy}\footnote{\url{https://github.com/fermiPy/fermipy}}. 
The best-fit position was determined to be R.A.=$98.264^\circ\pm0.039^\circ$, Dec.=$5.895^\circ\pm0.037^\circ$, with a 95\% confidence error circle of $0.^\circ093$ (Figure \ref{TSmap}).
The gamma-ray emission of HESS J0632+057 could be well represented by a power-law (PL) model, leading to a TS value of 28.5, a spectral index of 2.40$\pm$0.16 and an energy flux of (5.5$\pm$1.6)$\times$ 10$^{-12}$ erg cm$^{-2}$ s$^{-1}$ in 0.1-300 GeV.
The spectral index is consistent with that reported in \cite{2017ApJ...846..169L}.
The TS and flux level are both lower (TS=63 and flux {(9.2$\pm$1.6)$\times$ 10$^{-12}$ erg cm$^{-2}$ s$^{-1}$ in \cite{2017ApJ...846..169L}) but within 2$\sigma$ error.
We also tested other models with spectral curvature, e.g. a PL with a super-exponential cutoff, and a Log Parabola model. 
However, no significant improvement was detected.
{To explore if there is long-term flux variation, we divided the data into two sections, one with the same observation coverage with \cite{2017ApJ...846..169L} and other one with the rest. 
The first section yields a TS value of 18, an energy flux of (4.6$\pm$1.7)$\times$ 10$^{-12}$ erg cm$^{-2}$ s$^{-1}$ with a spectral index of 2.26$\pm$0.22 in 0.1-300 GeV.
The second section yields a TS value of 11, an energy flux of (7.4$\pm$2.8)$\times$ 10$^{-12}$ erg cm$^{-2}$ s$^{-1}$ and a spectral index of 2.5$\pm$0.2 in 0.1-300 GeV.
No flux difference could be claimed between the two time sections.
Our analysis used different Galactic diffuse model from \cite{2017ApJ...846..169L} (“gll\_iem\_v06.fits”).
Using the same Galactic diffuse model and observation time (first section above) with \cite{2017ApJ...846..169L}, we derived a higher TS value of 22 and but consistent flux level of (4.73$\pm$1.31)$\times$ 10$^{-12}$ erg cm$^{-2}$ s$^{-1}$ if using the current Galactic diffuse model.
Thus, the diffuse model is probably not the main reason for the result difference from \cite{2017ApJ...846..169L}, which may because of different background source models.
However, the difference in flux is within 2$\sigma$ error.
}
The energy spectral distribution (SED) of HESS J0632+057 is shown in Figure \ref{0632sed} with best fitted PL model.
The SED indicates a spectral turn-over above $\sim$10 GeV and likely below 100 GeV, suggesting another component above 10 GeV which connects with the TeV spectrum \citep{2021ApJ...923..241A}.
We tested a broken PL model in the likelihood fitting to explore the possible component.
However, no significant improvement was achieved and we could not explicitly determine the spectral turn-over because of the low statistics.

\begin{figure}[h]
    \centering
    \includegraphics[scale=0.5]{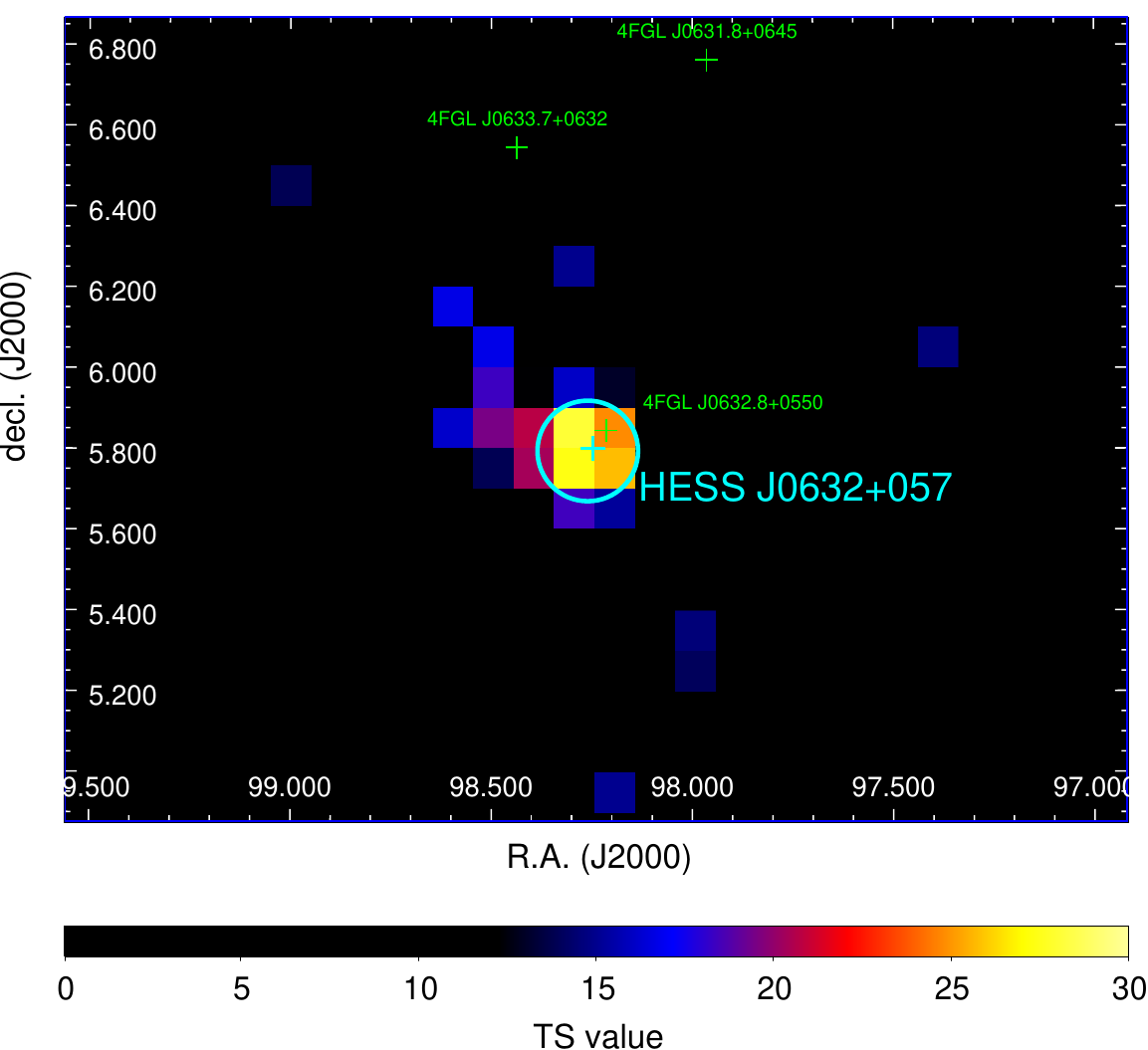}
    \caption{{\it Fermi}-LAT TS map of HESS J0632+057 region in 0.1-300 GeV.
The position of HESS J0632+057 is shown with a cyan cross and the 95\% confidence level of its gamma-ray emission is shown with a cyan circle.
Background 4FGL sources are shown with green crosses.
4FGL J0632.8+0550 is associated with HESS J0632+057 thus not included in the background model.
The x and y axes are R.A. and decl. (J2000, degrees).}
    \label{TSmap}
\end{figure}

\begin{figure}[h]
    \centering
    \includegraphics[scale=0.5]{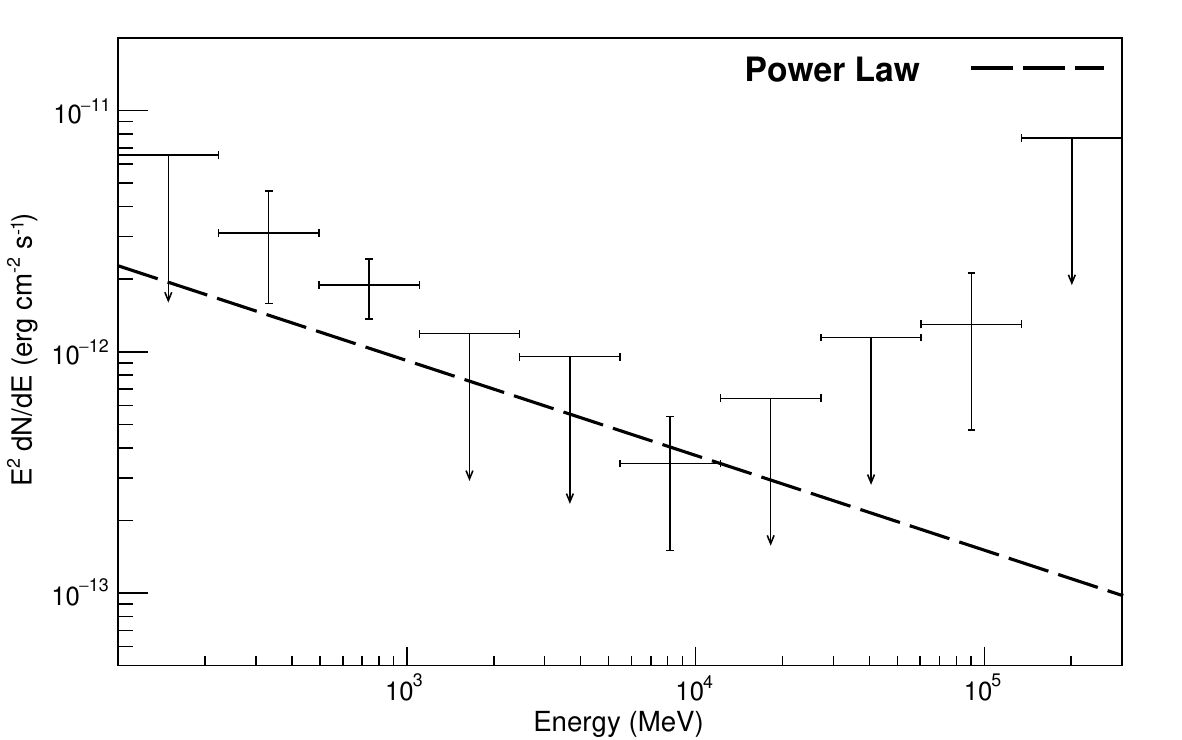}
    \caption{The SED and best fitted PL model of HESS J0632+057. A data point is calculated with TS$>$4 while a 95\% flux upper limit is derived with TS$<$4.     
    }
    \label{0632sed}
\end{figure}

HESS J0632+057 is identified by its long orbital period as a binary system with orbital-modulated multi-wavelength emissions. 
Its orbital characteristics have been extensively studied in the X-ray and TeV bands, supported by abundant data from various telescope observations.
In this study, we adopt an orbital period of 317.3 days and phase zero at MJD 54857.0, as determined from X-ray observations \citep{2021ApJ...923..241A}. 
The orbital phase-folded light curves in the X-ray (0.3–10 keV) and gamma-ray ($>$350 GeV) bands show an enhanced activity peak during phases 0.2–0.4 and a slightly lower peak around phases 0.6–0.8 \citep{2021ApJ...923..241A}.
To investigate whether the 0.1–300 GeV emission exhibits consistent orbital modulation with the X-ray and TeV, we divided the 15-year {\it Fermi}-LAT data into five equal orbital intervals with a binning of 0.2 orbital phase.
The orbital light curve of HESS J0632+057 is shown in \autoref{orbit_lc}.
The orbital light curve shows the highest TS value at phase 0.2-0.4, which aligns with the X-ray and TeV peaks.
However, because of limited statistics, we could not detect significant orbital modulation in 0.1-300 GeV for HESS J0632+057.

To explore the existence of GeV orbit-to-orbit variability as suggested in TeV range \citep{2021ApJ...923..241A}, we produced a long-term light curve of HESS J0632+057 in 0.1-300 GeV, adopting the binning of an orbital period, 317.3 days (\autoref{long_lc}).
However, because of limited statistics, no significant variability could be claimed.

\begin{figure}[h]
    \centering
    \includegraphics[width=0.7\linewidth]{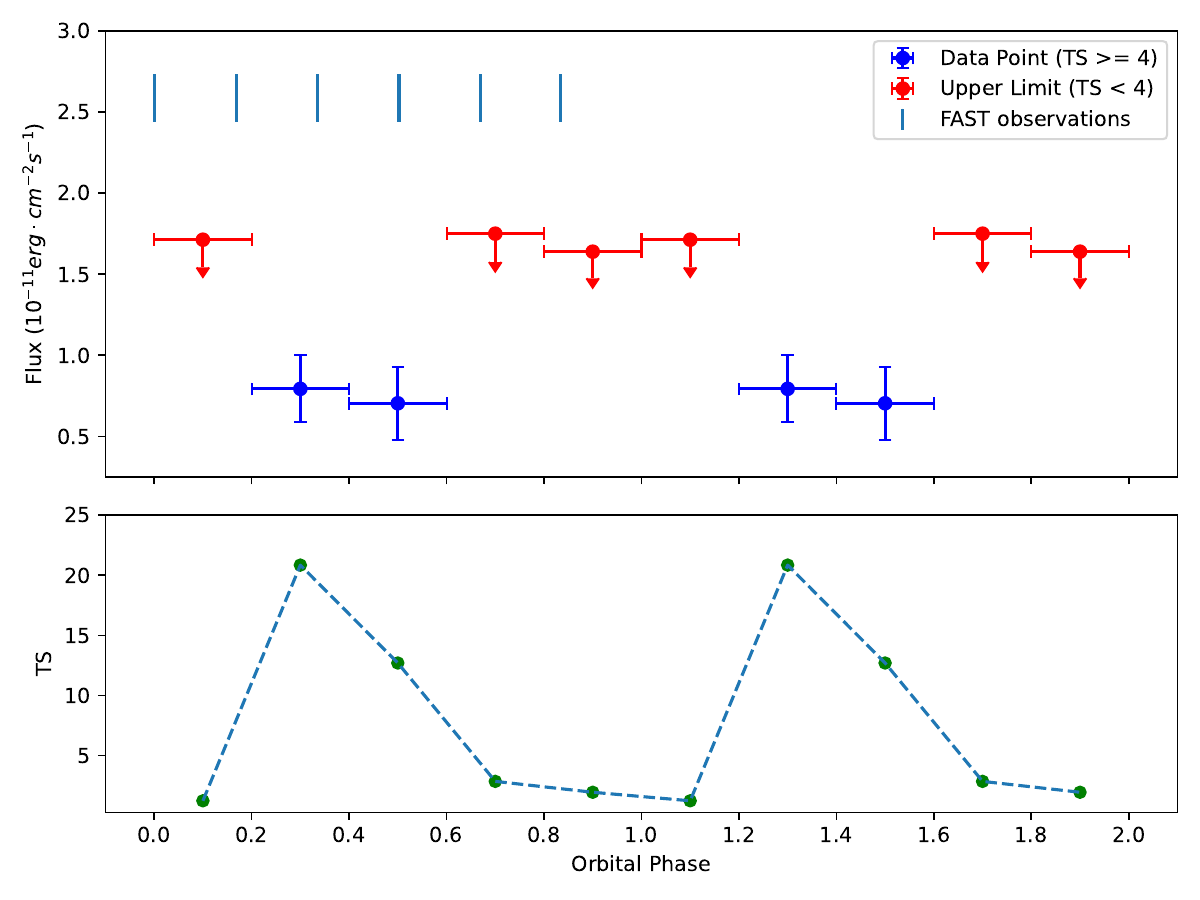}
    \caption{Orbital light curve (top) and TS value (bottom) of HESS J0632+057 in 0.1-300 GeV.}
    \label{orbit_lc}
\end{figure}

\begin{figure}[h]
    \centering
    \includegraphics[width=0.7\linewidth]{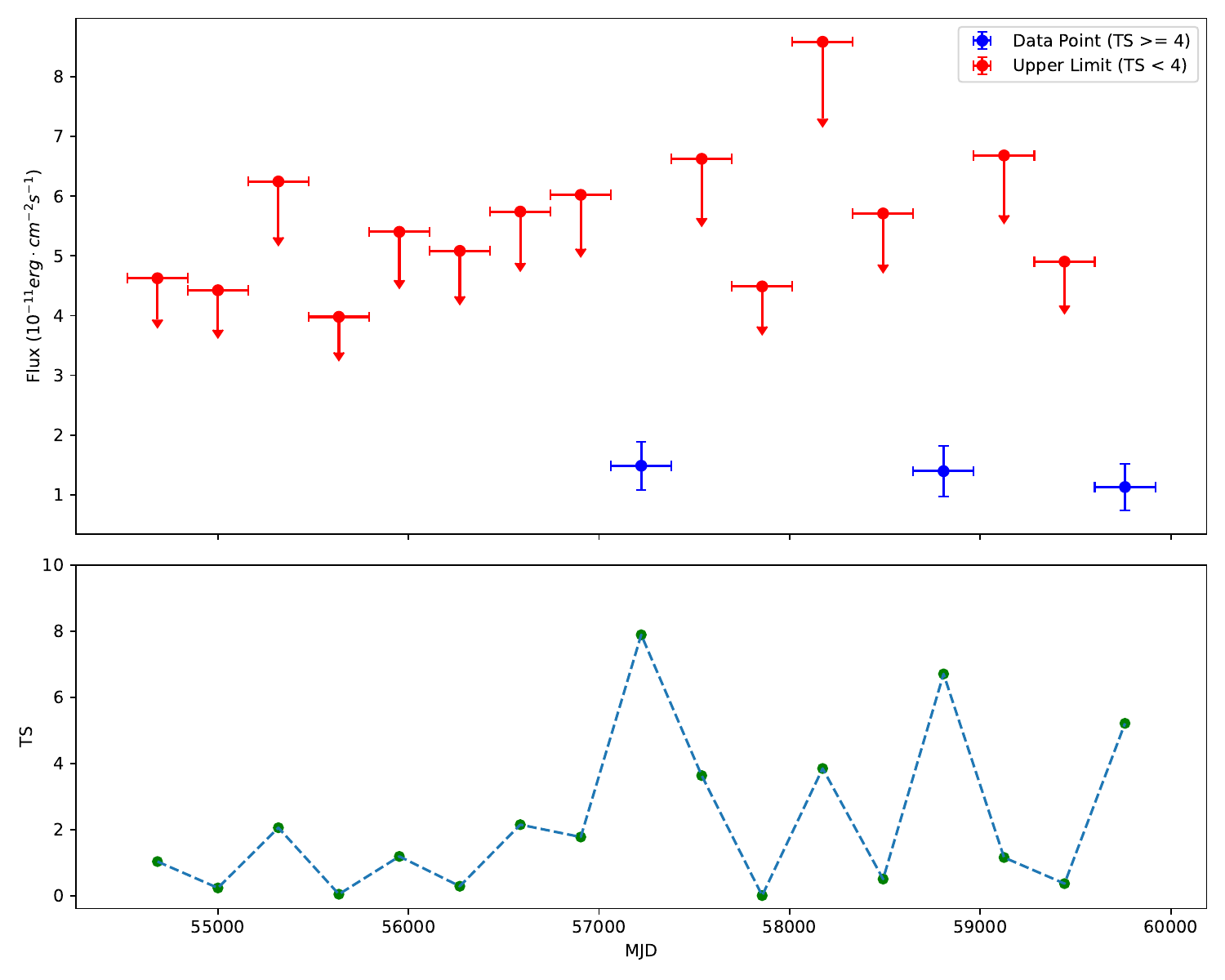}
    \caption{Long-term Light curve of HESS J0632+05 in 0.1-300 GeV with the binning of an orbit (317.3 days).}
    \label{long_lc}
\end{figure}

\clearpage

\section{Periodic signal search with FAST observations}
\label{pulse}
We carried out six deep radio observations of HESS J0632+057 by FAST \citep{2019SCPMA..6259502J,2020RAA....20...64J,2020Innov...100053Q} from September 2022 to May 2023 in 1.0-1.5 GHz to search for pulsations. 
The observation dates and durations are listed in \autoref{FAST:obs}. 
To assess the sensitivity of our search, we calculated the minimum detectable flux density
\begin{equation}
	S_{\rm min}=\frac{\beta(T_{\rm sys}+T_{\rm sky}){\rm SNR}}{{{\rm CF}_{\delta}\epsilon G\sqrt{n_{\rm p}t_{\rm obs}\Delta f}}}\sqrt{\frac{W}{P-W}}
 \label{eq:S}
\end{equation}
where $P$ is the period of the pulsar signal, $\beta\approx 1.05$ represents data digitization losses for our 8-bit recording system, $T_{\rm sys}+T_{\rm sky}=29\ {\rm K}$ is the sum of the telescope system temperature and the sky background temperature, $G=16 \ \rm K\ Jy^{-1}$ is the telescope gain, SNR denotes the signal-to-noise ratio of the pulsar signal, $n_{\rm p}=2$ is the number of polarization channels for data merging, $\Delta f=500\ {\rm MHz}$ is the frequency bandwidth for data merging, and $t_{\rm obs}$ is the observation duration. 
$\epsilon=0.93$ is the FFA Fast-folding Algorithm (FFA) search efficiency \citep{FFA_2020MNRAS.497.4654M}. $W$ refers to the width of the observed pulsar signal, encompassing both the intrinsic width of the pulsar signal and the broadening due to sampling time, scattering, and channel dispersion. $\rm CF_{\delta}$ accounts for the potential attenuation of the signal due to the removal of red noise \citep{2024SCPMA..6769512Z}. In our search, we set the running median filter window width for red noise removal to 2 seconds, meaning that pulse signals with widths less than 1 second would not be affected. Assuming a duty cycle of 10\%, the pulse width for a 10-second period would be 1 second, thus the removal of red noise from the data would have no effect on the assumed signal, hence ${\rm CF}_{\delta}\equiv 1$ in the equation. \autoref{FAST:obs} lists the FFA search sensitivity calculated using \autoref{eq:S} for a threshold SNR of 6, a signal duty cycle of 10\%, and a period range of 0.1-10 seconds.
\begin{table*}[h]
\centering
\caption{FAST observation information on HESS J0632+057.}
\label{FAST:obs}
\begin{threeparttable}
               \begin{tabular}{cccc}
            \toprule
            Observation date (UT)  &  \makecell{Observation \\length (s)} & \makecell{Corresponding \\orbital phase} &Sensitivity ($\mu$Jy)\tnote{*}\\
            \midrule
                  2022/09/06 & 4404 &0.50235-0.50250& 2\\
                    2022/10/29 & 4404 &0.33607-0.33623&2 \\
                    2022/12/21 & 4404 &0.16934-0.16949&2 \\
                    2023/02/12 & 4404 &0.99991-1.00006&2 \\
                    2023/04/06 & 4404 &0.83302-0.83317&2 \\
                    2023/05/29 & 4404 &0.66951-0.66967&2 \\
            \bottomrule
        \end{tabular}
    \begin{tablenotes}
        \item[*] The Sensitivity is the minimum detectable flux density assuming a duty cycle of 10\%, SNR of 6, and integration time for the corresponding observation length using FAST L-band receiver.
    \end{tablenotes}
\end{threeparttable}
\end{table*}

We first used \emph{presto}\footnote{\url{https://github.com/scottransom/presto}} to remove radio frequency interference (RFI) from the raw PSRFITS data. The data are then de-dispersed with a step size of 0.1 $\rm pc/cm^3$ in the range of 0-1000 $\rm pc/cm^3$, resulting in 10,000 time series corresponding to different dispersion measures. We employed an implementation of the fast-folding algorithm\footnote{\url{https://github.com/v-morello/riptide}} to search for periods in the range of 0.1-10 seconds in these time series, conducting searches over dispersion measure, period, and pulse width. Specifically, the time series for each dispersion measure are folded at different trial periods to form a profile within one period, and then box-car matched filtering with various widths is used to calculate the SNR of the profile. Once all dispersion measures, trial periods, and widths searches are completed, candidates with the highest SNR greater than 6 during the search process are marked for further visual inspection in both the frequency and time domains to determine if they are real signals. 
Our search resulted in tens of thousands of candidates across six observations. 
All of these candidates were visually inspected, but regrettably no authentic signals were detected. 
This indicates that no potential signal exceeds the minimum detectable flux listed in Table \ref{FAST:obs}.

\section{Summary \& Discussion}
\label{conclusion}

In this paper, we analyzed 15 years of {\it Fermi}-LAT data on HESS J0632+057, which is much longer than the 9 years of data presented in \cite{2017ApJ...846..169L}. 
Our results yielded a lower flux level compared to \cite{2017ApJ...846..169L}, which may because of
different background source model.
We explored orbit-to-orbit variability but could not claim it because of limited statistics.
Assuming a distance of 1.4 kpc \citep{2010ApJ...724..306A,2012MNRAS.421.1103C}, the luminosity of HESS J0632+057 in the 0.1–300 GeV band is estimated to be $\sim1.3 \times 10^{33} \mathrm{erg\cdot s^{-1}}$, which is among the lowest gamma-ray binaries \citep{2012Sci...335..189F, 2012ApJ...749...54H, 2013MNRAS.436..740C, 2015ApJ...811...68C, 2016ApJ...829..105C}.
The GeV orbital light curve suggests a consistent peak with X-ray and TeV \citep{2021ApJ...923..241A}, indicating the origin of a common particle population. 
Previous studies have shown that the hard X-ray and soft TeV spectra of HESS J0632+057 can be effectively modeled using a simple leptonic scenario \citep{2009ApJ...690L.101H,2012ApJ...754L..10A}. 
This indicates that X-rays originate from synchrotron radiation by high-energy particles, with TeV emissions produced via inverse Compton (IC) scattering with the companion star's photon field.
Our GeV SED hints a spectral turn-over at $\sim$10-100 GeV, and another component above it which connects with the TeV spectrum \citep{2021ApJ...923..241A}. 
This is consistent with \cite{2013MNRAS.436..740C} which proposed a spectral turn-over in the LAT range and \cite{2017ApJ...846..169L} that suggests the turn-over to be above 10 GeV.
However, we could not explicitly locate the spectral turn-over because of limited statistics.
The possible spectral turn-over could be the result of pair production on stellar photons for gamma-rays above $\sim$50 GeV, or another emission component in the TeV range.

We carried out six deep FAST observations on HESS J0632+057, evenly distributed in the orbital phase (Table 1).
Periodic pulsations were searched within these observations but no signal was detected.
Assuming a signal duty cycle of 10\%, and a period range of 0.1-10 seconds, we calculated a search sensitivity of 2 $\mu$Jy.
Radio pulsations have been detected from gamma-ray binary PSR B1259-63 \citep{1992ApJ...387L..37J}, PSR J2032+4127 \citep{2009ApJ...705....1C} and LSI +61 303 \citep{2022NatAs...6..698W}.
Radio pulsation of LSI +61 303 was revealed by FAST with a flux level of 4.40 $\mu$Jy which is above the sensitivity of our FAST observations.
Considering the non-steady nature of the radio pulsation in LSI +61 303, the non-detention of periodic signal in HESS J0632+057 does not rule out the existence of a pulsar.
As pointed out by \cite{2022NatAs...6..698W}, changes in the interstellar scintillation, pulsar nulling, and variations in stellar wind, could all lead to a transient behavior of the pulsation and thus be missed with our FAST observations.
Additionally, the non-detection of radio pulsation may also be attributed to other reasons. 
The radio beam of the pulsar may not be pointing towards Earth.
Assuming HESS J0632+057 hosts a pulsar similar to PSR B1259-63 (period $P=0.0477625$s, period derivative $\dot{P}=2.27875\times10^{-15}$ s s$^{-1}$), following \cite{Emmering1989ApJ...345..931E,Kijak2003A&A...397..969K,Takata2011ApJ...726...44T,chen2021A&A...652A..39C} the radio beaming fraction at 1.25 GHz would be $\sim$ 50$\%$, which is comparably large.
However, even if the radio beam is pointing towards us, the free-free absorption in stellar wind of MWC 148 may lead to the non-detection.
If we assume a wind velocity of 1000 km s$^{-1}$, temperature of $10^{4}$k\cite{waters1988A&A...198..200W}, distance between pulsar and massive star d$\sim$4 AU\cite{2025MNRAS.536..166M}, Gaunt factor $g_{ff}\sim1$\cite{chen2021A&A...652A..39C},
the non-detection of radio pulsation (free-free optical depth $\tau_{ff}\geq1$) at 1.25 GHz would lead to a stellar wind mass loss rate $\dot{M}\geq5\times10^{-8} M_\odot\ yr^{-1}$, adopting equation (2) of \cite{2013A&ARv..21...64D}, which is consistent with the value adopted in \cite{wind_2020ApJ...888..115A}.
The companion star MWC 148 has a similar mass and radius as that in LS I+61 303 but is five times bigger in the size of the circumstellar disc \citep{2016A&A...593A..97Z}.
The stellar disk inclination angle is estimated as $i_d\approx12^\circ$ by \cite{2022A&A...658A.153C} and is almost co-planar with the orbital plane, which may dominate the absorption of radio pulsation.
Additionally, extended radio emission has been detected from HESS J0632+057\cite{extended_radio_2011A&A...533L...7M} which likely originates from the shock between stellar wind and pulsar wind.
It may overshine the radio pulsation and bring further difficulty to the detection.
Future radio observations with denser orbital phase coverage may test above possibilities and lead to the pulsar detection in HESS J0632+057.

\clearpage
\textit{Acknowledgments}

This work is supported by National Natural Science Foundation of China (NSFC) Programs No.12273038.
DFT acknowledges support from the Spanish grants PID2021-124581OB-I00, 2021SGR00426, CEX2020-001058-M and EU PRTR-C17.I1.
We acknowledge the assistance from Dr. M. Kerr on the gamma-ray ephemeris for PSR J0633+0632.
The \textit{Fermi} LAT Collaboration acknowledges generous ongoing support
from a number of agencies and institutes that have supported both the
development and the operation of the LAT as well as scientific data analysis.
These include the National Aeronautics and Space Administration and the
Department of Energy in the United States, the Commissariat \`a l'Energie Atomique
and the Centre National de la Recherche Scientifique / Institut National de Physique
Nucl\'eaire et de Physique des Particules in France, the Agenzia Spaziale Italiana
and the Istituto Nazionale di Fisica Nucleare in Italy, the Ministry of Education,
Culture, Sports, Science and Technology (MEXT), High Energy Accelerator Research
Organization (KEK) and Japan Aerospace Exploration Agency (JAXA) in Japan, and
the K.~A.~Wallenberg Foundation, the Swedish Research Council and the
Swedish National Space Board in Sweden.
Additional support for science analysis during the operations phase is gratefully
acknowledged from the Istituto Nazionale di Astrofisica in Italy and the Centre
National d'\'Etudes Spatiales in France. This work performed in part under DOE
Contract DE-AC02-76SF00515.


\bibliographystyle{elsarticle-num} 
\bibliography{sample631}

\begin{thebibliography}{10}
\expandafter\ifx\csname url\endcsname\relax
  \def\url#1{\texttt{#1}}\fi
\expandafter\ifx\csname urlprefix\endcsname\relax\def\urlprefix{URL }\fi
\expandafter\ifx\csname href\endcsname\relax
  \def\href#1#2{#2} \def\path#1{#1}\fi

\bibitem{2013A&ARv..21...64D}
G.~{Dubus}, {Gamma-ray binaries and related systems}, A\&ARv 21 (2013) 64.
\newblock \href {http://arxiv.org/abs/1307.7083} {\path{arXiv:1307.7083}}, \href {https://doi.org/10.1007/s00159-013-0064-5} {\path{doi:10.1007/s00159-013-0064-5}}.

\bibitem{2005Sci...309..746A}
F.~{Aharonian}, A.~G. {Akhperjanian}, K.~M. {Aye}, A.~R. {Bazer-Bachi}, M.~{Beilicke}, W.~{Benbow}, D.~{Berge}, P.~{Berghaus}, K.~{Bernl{\"o}hr}, C.~{Boisson}, O.~{Bolz}, V.~{Borrel}, I.~{Braun}, F.~{Breitling}, A.~M. {Brown}, J.~B. {Gordo}, P.~M. {Chadwick}, L.~M. {Chounet}, R.~{Cornils}, L.~{Costamante}, B.~{Degrange}, H.~J. {Dickinson}, A.~{Djannati-Ata{\"\i}}, L.~O. {Drury}, G.~{Dubus}, D.~{Emmanoulopoulos}, P.~{Espigat}, F.~{Feinstein}, P.~{Fleury}, G.~{Fontaine}, Y.~{Fuchs}, S.~{Funk}, Y.~A. {Gallant}, B.~{Giebels}, S.~{Gillessen}, J.~F. {Glicenstein}, P.~{Goret}, C.~{Hadjichristidis}, M.~{Hauser}, G.~{Heinzelmann}, G.~{Henri}, G.~{Hermann}, J.~A. {Hinton}, W.~{Hofmann}, M.~{Holleran}, D.~{Horns}, A.~{Jacholkowska}, O.~C. {de Jager}, B.~{Kh{\'e}lifi}, N.~{Komin}, A.~{Konopelko}, I.~J. {Latham}, R.~{Le Gallou}, A.~{Lemi{\`e}re}, M.~{Lemoine-Goumard}, N.~{Leroy}, T.~{Lohse}, A.~{Marcowith}, J.~M. {Martin}, O.~{Martineau-Huynh}, C.~{Masterson}, T.~J.~L. {McComb}, M.~{de Naurois}, S.~J. {Nolan},
  A.~{Noutsos}, K.~J. {Orford}, J.~L. {Osborne}, M.~{Ouchrif}, M.~{Panter}, G.~{Pelletier}, S.~{Pita}, G.~{P{\"u}hlhofer}, M.~{Punch}, B.~C. {Raubenheimer}, M.~{Raue}, J.~{Raux}, S.~M. {Rayner}, A.~{Reimer}, O.~{Reimer}, J.~{Ripken}, L.~{Rob}, L.~{Rolland}, G.~{Rowell}, V.~{Sahakian}, L.~{Saug{\'e}}, S.~{Schlenker}, R.~{Schlickeiser}, C.~{Schuster}, U.~{Schwanke}, M.~{Siewert}, H.~{Sol}, D.~{Spangler}, R.~{Steenkamp}, C.~{Stegmann}, J.~P. {Tavernet}, R.~{Terrier}, C.~G. {Th{\'e}oret}, M.~{Tluczykont}, G.~{Vasileiadis}, C.~{Venter}, P.~{Vincent}, H.~J. {V{\"o}lk}, S.~J. {Wagner}, {Discovery of Very High Energy Gamma Rays Associated with an X-ray Binary}, Science 309~(5735) (2005) 746--749.
\newblock \href {http://arxiv.org/abs/astro-ph/0508298} {\path{arXiv:astro-ph/0508298}}, \href {https://doi.org/10.1126/science.1113764} {\path{doi:10.1126/science.1113764}}.

\bibitem{2006A&A...460..743A}
F.~{Aharonian}, A.~G. {Akhperjanian}, A.~R. {Bazer-Bachi}, M.~{Beilicke}, W.~{Benbow}, D.~{Berge}, K.~{Bernl{\"o}hr}, C.~{Boisson}, O.~{Bolz}, V.~{Borrel}, I.~{Braun}, A.~M. {Brown}, R.~{B{\"u}hler}, I.~{B{\"u}sching}, S.~{Carrigan}, P.~M. {Chadwick}, L.~M. {Chounet}, R.~{Cornils}, L.~{Costamante}, B.~{Degrange}, H.~J. {Dickinson}, A.~{Djannati-Ata{\"\i}}, L.~{O'C. Drury}, G.~{Dubus}, K.~{Egberts}, D.~{Emmanoulopoulos}, P.~{Espigat}, F.~{Feinstein}, E.~{Ferrero}, A.~{Fiasson}, G.~{Fontaine}, S.~{Funk}, S.~{Funk}, M.~{F{\"u}{\ss}ling}, Y.~A. {Gallant}, B.~{Giebels}, J.~F. {Glicenstein}, P.~{Goret}, C.~{Hadjichristidis}, D.~{Hauser}, M.~{Hauser}, G.~{Heinzelmann}, G.~{Henri}, G.~{Hermann}, J.~A. {Hinton}, A.~{Hoffmann}, W.~{Hofmann}, M.~{Holleran}, D.~{Horns}, A.~{Jacholkowska}, O.~C. {de Jager}, E.~{Kendziorra}, B.~{Kh{\'e}lifi}, N.~{Komin}, A.~{Konopelko}, K.~{Kosack}, I.~J. {Latham}, R.~{Le Gallou}, A.~{Lemi{\`e}re}, M.~{Lemoine-Goumard}, T.~{Lohse}, J.~M. {Martin}, O.~{Martineau-Huynh}, A.~{Marcowith},
  C.~{Masterson}, G.~{Maurin}, T.~J.~L. {McComb}, E.~{Moulin}, M.~{de Naurois}, D.~{Nedbal}, S.~J. {Nolan}, A.~{Noutsos}, K.~J. {Orford}, J.~L. {Osborne}, M.~{Ouchrif}, M.~{Panter}, G.~{Pelletier}, S.~{Pita}, G.~{P{\"u}hlhofer}, M.~{Punch}, B.~C. {Raubenheimer}, M.~{Raue}, S.~M. {Rayner}, A.~{Reimer}, O.~{Reimer}, J.~{Ripken}, L.~{Rob}, L.~{Rolland}, G.~{Rowell}, V.~{Sahakian}, A.~{Santangelo}, L.~{Saug{\'e}}, S.~{Schlenker}, R.~{Schlickeiser}, R.~{Schr{\"o}der}, U.~{Schwanke}, S.~{Schwarzburg}, A.~{Shalchi}, H.~{Sol}, D.~{Spangler}, F.~{Spanier}, R.~{Steenkamp}, C.~{Stegmann}, G.~{Superina}, J.~P. {Tavernet}, R.~{Terrier}, M.~{Tluczykont}, C.~{van Eldik}, G.~{Vasileiadis}, C.~{Venter}, P.~{Vincent}, H.~J. {V{\"o}lk}, S.~J. {Wagner}, M.~{Ward}, {3.9 day orbital modulation in the TeV {\ensuremath{\gamma}}-ray flux and spectrum from the X-ray binary LS 5039}, A\&A 460~(3) (2006) 743--749.
\newblock \href {http://arxiv.org/abs/astro-ph/0607192} {\path{arXiv:astro-ph/0607192}}, \href {https://doi.org/10.1051/0004-6361:20065940} {\path{doi:10.1051/0004-6361:20065940}}.

\bibitem{2009ApJ...701L.123A}
A.~A. {Abdo}, M.~{Ackermann}, M.~{Ajello}, W.~B. {Atwood}, M.~{Axelsson}, L.~{Baldini}, J.~{Ballet}, G.~{Barbiellini}, D.~{Bastieri}, B.~M. {Baughman}, K.~{Bechtol}, R.~{Bellazzini}, B.~{Berenji}, R.~{Blandford}, E.~D. {Bloom}, E.~{Bonamente}, A.~W. {Borgland}, J.~{Bregeon}, A.~{Brez}, M.~{Brigida}, P.~{Bruel}, T.~H. {Burnett}, G.~A. {Caliandro}, R.~A. {Cameron}, P.~A. {Caraveo}, J.~M. {Casandjian}, E.~{Cavazzuti}, C.~{Cecchi}, {\"O}.~{{\c{C}}elik}, E.~{Charles}, S.~{Chaty}, A.~{Chekhtman}, C.~C. {Cheung}, J.~{Chiang}, S.~{Ciprini}, R.~{Claus}, J.~{Cohen-Tanugi}, L.~R. {Cominsky}, J.~{Conrad}, S.~{Corbel}, R.~{Corbet}, S.~{Cutini}, C.~D. {Dermer}, A.~{de Angelis}, A.~{de Luca}, F.~{de Palma}, S.~W. {Digel}, M.~{Dormody}, E.~{do Couto e Silva}, P.~S. {Drell}, R.~{Dubois}, G.~{Dubus}, D.~{Dumora}, C.~{Farnier}, C.~{Favuzzi}, S.~J. {Fegan}, W.~B. {Focke}, M.~{Frailis}, Y.~{Fukazawa}, S.~{Funk}, P.~{Fusco}, F.~{Gargano}, D.~{Gasparrini}, N.~{Gehrels}, S.~{Germani}, B.~{Giebels}, N.~{Giglietto}, F.~{Giordano},
  T.~{Glanzman}, G.~{Godfrey}, I.~A. {Grenier}, M.~H. {Grondin}, J.~E. {Grove}, L.~{Guillemot}, S.~{Guiriec}, Y.~{Hanabata}, A.~K. {Harding}, M.~{Hayashida}, E.~{Hays}, A.~B. {Hill}, R.~E. {Hughes}, G.~{J{\'o}hannesson}, A.~S. {Johnson}, R.~P. {Johnson}, T.~J. {Johnson}, W.~N. {Johnson}, T.~{Kamae}, H.~{Katagiri}, J.~{Kataoka}, N.~{Kawai}, M.~{Kerr}, J.~{Kn{\"o}dlseder}, M.~L. {Kocian}, F.~{Kuehn}, M.~{Kuss}, J.~{Lande}, S.~{Larsson}, L.~{Latronico}, F.~{Longo}, F.~{Loparco}, B.~{Lott}, M.~N. {Lovellette}, P.~{Lubrano}, G.~M. {Madejski}, A.~{Makeev}, M.~{Marelli}, M.~N. {Mazziotta}, J.~E. {McEnery}, C.~{Meurer}, P.~F. {Michelson}, W.~{Mitthumsiri}, T.~{Mizuno}, C.~{Monte}, M.~E. {Monzani}, A.~{Morselli}, I.~V. {Moskalenko}, S.~{Murgia}, P.~L. {Nolan}, E.~{Nuss}, T.~{Ohsugi}, A.~{Okumura}, N.~{Omodei}, E.~{Orlando}, J.~F. {Ormes}, D.~{Paneque}, J.~H. {Panetta}, D.~{Parent}, V.~{Pelassa}, M.~{Pepe}, M.~{Pesce-Rollins}, F.~{Piron}, T.~A. {Porter}, S.~{Rain{\`o}}, R.~{Rando}, P.~S. {Ray}, M.~{Razzano}, N.~{Rea},
  A.~{Reimer}, O.~{Reimer}, T.~{Reposeur}, S.~{Ritz}, L.~S. {Rochester}, A.~Y. {Rodriguez}, R.~W. {Romani}, F.~{Ryde}, H.~F.~W. {Sadrozinski}, D.~{Sanchez}, A.~{Sander}, P.~M. {Saz Parkinson}, J.~D. {Scargle}, C.~{Sgr{\`o}}, M.~S. {Shaw}, A.~{Sierpowska-Bartosik}, E.~J. {Siskind}, D.~A. {Smith}, P.~D. {Smith}, G.~{Spandre}, P.~{Spinelli}, E.~{Striani}, M.~S. {Strickman}, D.~J. {Suson}, H.~{Tajima}, H.~{Takahashi}, T.~{Takahashi}, T.~{Tanaka}, J.~B. {Thayer}, J.~G. {Thayer}, D.~J. {Thompson}, L.~{Tibaldo}, D.~F. {Torres}, G.~{Tosti}, A.~{Tramacere}, Y.~{Uchiyama}, T.~L. {Usher}, V.~{Vasileiou}, N.~{Vilchez}, V.~{Vitale}, A.~P. {Waite}, P.~{Wang}, B.~L. {Winer}, K.~S. {Wood}, T.~{Ylinen}, M.~{Ziegler}, {Fermi LAT Observations of LS I +61{\textdegree}303: First Detection of an Orbital Modulation in GeV Gamma Rays}, ApJL 701~(2) (2009) L123--L128.
\newblock \href {http://arxiv.org/abs/0907.4307} {\path{arXiv:0907.4307}}, \href {https://doi.org/10.1088/0004-637X/701/2/L123} {\path{doi:10.1088/0004-637X/701/2/L123}}.

\bibitem{2005A&A...442....1A}
F.~{Aharonian}, A.~G. {Akhperjanian}, K.~M. {Aye}, A.~R. {Bazer-Bachi}, M.~{Beilicke}, W.~{Benbow}, D.~{Berge}, P.~{Berghaus}, K.~{Bernl{\"o}hr}, C.~{Boisson}, O.~{Bolz}, I.~{Braun}, F.~{Breitling}, A.~M. {Brown}, J.~{Bussons Gordo}, P.~M. {Chadwick}, L.~M. {Chounet}, R.~{Cornils}, L.~{Costamante}, B.~{Degrange}, A.~{Djannati-Ata{\"\i}}, L.~{O'C. Drury}, G.~{Dubus}, D.~{Emmanoulopoulos}, P.~{Espigat}, F.~{Feinstein}, P.~{Fleury}, G.~{Fontaine}, Y.~{Fuchs}, S.~{Funk}, Y.~A. {Gallant}, B.~{Giebels}, S.~{Gillessen}, J.~F. {Glicenstein}, P.~{Goret}, C.~{Hadjichristidis}, M.~{Hauser}, G.~{Heinzelmann}, G.~{Henri}, G.~{Hermann}, J.~A. {Hinton}, W.~{Hofmann}, M.~{Holleran}, D.~{Horns}, O.~C. {de Jager}, S.~{Johnston}, B.~{Kh{\'e}lifi}, J.~G. {Kirk}, N.~{Komin}, A.~{Konopelko}, I.~J. {Latham}, R.~{Le Gallou}, A.~{Lemi{\`e}re}, M.~{Lemoine-Goumard}, N.~{Leroy}, O.~{Martineau-Huynh}, T.~{Lohse}, A.~{Marcowith}, C.~{Masterson}, T.~J.~L. {McComb}, M.~{de Naurois}, S.~J. {Nolan}, A.~{Noutsos}, K.~J. {Orford}, J.~L.
  {Osborne}, M.~{Ouchrif}, M.~{Panter}, G.~{Pelletier}, S.~{Pita}, G.~{P{\"u}hlhofer}, M.~{Punch}, B.~C. {Raubenheimer}, M.~{Raue}, J.~{Raux}, S.~M. {Rayner}, I.~{Redondo}, A.~{Reimer}, O.~{Reimer}, J.~{Ripken}, L.~{Rob}, L.~{Rolland}, G.~{Rowell}, V.~{Sahakian}, L.~{Saug{\'e}}, S.~{Schlenker}, R.~{Schlickeiser}, C.~{Schuster}, U.~{Schwanke}, M.~{Siewert}, O.~{Skj{\ae}raasen}, H.~{Sol}, R.~{Steenkamp}, C.~{Stegmann}, J.~P. {Tavernet}, R.~{Terrier}, C.~G. {Th{\'e}oret}, M.~{Tluczykont}, G.~{Vasileiadis}, C.~{Venter}, P.~{Vincent}, H.~J. {V{\"o}lk}, S.~J. {Wagner}, {Discovery of the binary pulsar PSR B1259-63 in very-high-energy gamma rays around periastron with HESS}, A\&A 442~(1) (2005) 1--10.
\newblock \href {http://arxiv.org/abs/astro-ph/0506280} {\path{arXiv:astro-ph/0506280}}, \href {https://doi.org/10.1051/0004-6361:20052983} {\path{doi:10.1051/0004-6361:20052983}}.

\bibitem{2012Sci...335..189F}
{Fermi LAT Collaboration}, M.~{Ackermann}, M.~{Ajello}, J.~{Ballet}, G.~{Barbiellini}, D.~{Bastieri}, A.~{Belfiore}, R.~{Bellazzini}, B.~{Berenji}, R.~D. {Blandford}, E.~D. {Bloom}, E.~{Bonamente}, A.~W. {Borgland}, J.~{Bregeon}, M.~{Brigida}, P.~{Bruel}, R.~{Buehler}, S.~{Buson}, G.~A. {Caliandro}, R.~A. {Cameron}, P.~A. {Caraveo}, E.~{Cavazzuti}, C.~{Cecchi}, {\"O}.~{{\c{C}}elik}, E.~{Charles}, S.~{Chaty}, A.~{Chekhtman}, C.~C. {Cheung}, J.~{Chiang}, {Ciprini}, S.~{}, R.~{Claus}, J.~{Cohen-Tanugi}, S.~{Corbel}, R.~H.~D. {Corbet}, S.~{Cutini}, A.~{de Luca}, P.~R. {den Hartog}, F.~{de Palma}, C.~D. {Dermer}, S.~W. {Digel}, E.~{do Couto e Silva}, D.~{Donato}, P.~S. {Drell}, A.~{Drlica-Wagner}, R.~{Dubois}, G.~{Dubus}, C.~{Favuzzi}, S.~J. {Fegan}, E.~C. {Ferrara}, W.~B. {Focke}, P.~{Fortin}, Y.~{Fukazawa}, S.~{Funk}, P.~{Fusco}, F.~{Gargano}, D.~{Gasparrini}, N.~{Gehrels}, S.~{Germani}, N.~{Giglietto}, F.~{Giordano}, M.~{Giroletti}, T.~{Glanzman}, G.~{Godfrey}, I.~A. {Grenier}, J.~E. {Grove}, S.~{Guiriec},
  D.~{Hadasch}, Y.~{Hanabata}, A.~K. {Harding}, M.~{Hayashida}, E.~{Hays}, A.~B. {Hill}, R.~E. {Hughes}, G.~{J{\'o}hannesson}, A.~S. {Johnson}, T.~J. {Johnson}, T.~{Kamae}, H.~{Katagiri}, J.~{Kataoka}, M.~{Kerr}, J.~{Kn{\"o}dlseder}, M.~{Kuss}, J.~{Lande}, F.~{Longo}, F.~{Loparco}, M.~N. {Lovellette}, P.~{Lubrano}, M.~N. {Mazziotta}, J.~E. {McEnery}, P.~F. {Michelson}, W.~{Mitthumsiri}, T.~{Mizuno}, C.~{Monte}, M.~E. {Monzani}, A.~{Morselli}, I.~V. {Moskalenko}, S.~{Murgia}, T.~{Nakamori}, M.~{Naumann-Godo}, J.~P. {Norris}, E.~{Nuss}, M.~{Ohno}, T.~{Ohsugi}, A.~{Okumura}, N.~{Omodei}, E.~{Orlando}, M.~{Ozaki}, D.~{Paneque}, D.~{Parent}, M.~{Pesce-Rollins}, M.~{Pierbattista}, F.~{Piron}, G.~{Pivato}, T.~A. {Porter}, S.~{Rain{\`o}}, R.~{Rando}, M.~{Razzano}, A.~{Reimer}, O.~{Reimer}, S.~{Ritz}, R.~W. {Romani}, M.~{Roth}, P.~M. {Saz Parkinson}, C.~{Sgr{\`o}}, E.~J. {Siskind}, G.~{Spandre}, P.~{Spinelli}, D.~J. {Suson}, H.~{Takahashi}, T.~{Tanaka}, J.~G. {Thayer}, J.~B. {Thayer}, D.~J. {Thompson}, L.~{Tibaldo},
  M.~{Tinivella}, D.~F. {Torres}, G.~{Tosti}, E.~{Troja}, Y.~{Uchiyama}, T.~L. {Usher}, J.~{Vandenbroucke}, G.~{Vianello}, V.~{Vitale}, A.~P. {Waite}, B.~L. {Winer}, K.~S. {Wood}, M.~{Wood}, Z.~{Yang}, S.~{Zimmer}, M.~J. {Coe}, F.~{Di Mille}, P.~G. {Edwards}, M.~D. {Filipovi{\'c}}, J.~L. {Payne}, J.~{Stevens}, M.~A.~P. {Torres}, {Periodic Emission from the Gamma-Ray Binary 1FGL J1018.6-5856}, Science 335~(6065) (2012) 189.
\newblock \href {http://arxiv.org/abs/1202.3164} {\path{arXiv:1202.3164}}, \href {https://doi.org/10.1126/science.1213974} {\path{doi:10.1126/science.1213974}}.

\bibitem{2019ApJ...884...93C}
R.~H.~D. {Corbet}, L.~{Chomiuk}, M.~J. {Coe}, J.~B. {Coley}, G.~{Dubus}, P.~G. {Edwards}, P.~{Martin}, V.~A. {McBride}, J.~{Stevens}, J.~{Strader}, L.~J. {Townsend}, {Discovery of the Galactic High-mass Gamma-Ray Binary 4FGL J1405.1-6119}, ApJ 884~(1) (2019) 93.
\newblock \href {http://arxiv.org/abs/1908.10764} {\path{arXiv:1908.10764}}, \href {https://doi.org/10.3847/1538-4357/ab3e32} {\path{doi:10.3847/1538-4357/ab3e32}}.

\bibitem{2018ApJ...867L..19A}
A.~U. {Abeysekara}, W.~{Benbow}, R.~{Bird}, A.~{Brill}, R.~{Brose}, J.~H. {Buckley}, A.~J. {Chromey}, M.~K. {Daniel}, A.~{Falcone}, J.~P. {Finley}, L.~{Fortson}, A.~{Furniss}, A.~{Gent}, G.~H. {Gillanders}, D.~{Hanna}, T.~{Hassan}, O.~{Hervet}, J.~{Holder}, G.~{Hughes}, T.~B. {Humensky}, P.~{Kaaret}, P.~{Kar}, M.~{Kertzman}, D.~{Kieda}, M.~{Krause}, F.~{Krennrich}, S.~{Kumar}, M.~J. {Lang}, T.~T.~Y. {Lin}, G.~{Maier}, P.~{Moriarty}, R.~{Mukherjee}, S.~{O'Brien}, R.~A. {Ong}, A.~N. {Otte}, N.~{Park}, A.~{Petrashyk}, M.~{Pohl}, E.~{Pueschel}, J.~{Quinn}, K.~{Ragan}, G.~T. {Richards}, E.~{Roache}, I.~{Sadeh}, M.~{Santander}, S.~{Schlenstedt}, G.~H. {Sembroski}, I.~{Sushch}, J.~{Tyler}, V.~V. {Vassiliev}, S.~P. {Wakely}, A.~{Weinstein}, R.~M. {Wells}, P.~{Wilcox}, A.~{Wilhelm}, D.~A. {Williams}, T.~J. {Williamson}, B.~{Zitzer}, {VERITAS Collaboration}, V.~A. {Acciari}, S.~{Ansoldi}, L.~A. {Antonelli}, A.~{Arbet Engels}, D.~{Baack}, A.~{Babi{\'c}}, B.~{Banerjee}, U.~{Barres de Almeida}, J.~A. {Barrio}, J.~{Becerra
  Gonz{\'a}lez}, W.~{Bednarek}, E.~{Bernardini}, A.~{Berti}, J.~{Besenrieder}, W.~{Bhattacharyya}, C.~{Bigongiari}, A.~{Biland}, O.~{Blanch}, G.~{Bonnoli}, G.~{Busetto}, R.~{Carosi}, G.~{Ceribella}, S.~{Cikota}, S.~M. {Colak}, P.~{Colin}, E.~{Colombo}, J.~L. {Contreras}, J.~{Cortina}, S.~{Covino}, V.~{D'Elia}, P.~{Da Vela}, F.~{Dazzi}, A.~{De Angelis}, B.~{De Lotto}, M.~{Delfino}, J.~{Delgado}, F.~{Di Pierro}, E.~{Do Souto Espi{\~n}era}, A.~{Dom{\'\i}nguez}, D.~{Dominis Prester}, D.~{Dorner}, M.~{Doro}, S.~{Einecke}, D.~{Elsaesser}, V.~{Fallah Ramazani}, A.~{Fattorini}, A.~{Fern{\'a}ndez-Barral}, G.~{Ferrara}, D.~{Fidalgo}, L.~{Foffano}, M.~V. {Fonseca}, L.~{Font}, C.~{Fruck}, D.~{Galindo}, S.~{Gallozzi}, R.~J. {Garc{\'\i}a L{\'o}pez}, M.~{Garczarczyk}, S.~{Gasparyan}, M.~{Gaug}, P.~{Giammaria}, N.~{Godinovi{\'c}}, D.~{Guberman}, D.~{Hadasch}, A.~{Hahn}, J.~{Herrera}, J.~{Hoang}, D.~{Hrupec}, S.~{Inoue}, K.~{Ishio}, Y.~{Iwamura}, H.~{Kubo}, J.~{Kushida}, D.~{Kuve{\v{z}}di{\'c}}, A.~{Lamastra}, D.~{Lelas},
  F.~{Leone}, E.~{Lindfors}, S.~{Lombardi}, F.~{Longo}, M.~{L{\'o}pez}, A.~{L{\'o}pez-Oramas}, B.~{Machado de Oliveira Fraga}, C.~{Maggio}, P.~{Majumdar}, M.~{Makariev}, M.~{Mallamaci}, G.~{Maneva}, M.~{Manganaro}, K.~{Mannheim}, L.~{Maraschi}, M.~{Mariotti}, M.~{Mart{\'\i}nez}, S.~{Masuda}, D.~{Mazin}, M.~{Minev}, J.~M. {Miranda}, R.~{Mirzoyan}, E.~{Molina}, A.~{Moralejo}, V.~{Moreno}, E.~{Moretti}, P.~{Munar-Adrover}, V.~{Neustroev}, A.~{Niedzwiecki}, M.~{Nievas Rosillo}, C.~{Nigro}, K.~{Nilsson}, D.~{Ninci}, K.~{Nishijima}, K.~{Noda}, L.~{Nogu{\'e}s}, M.~{N{\"o}the}, S.~{Paiano}, J.~{Palacio}, D.~{Paneque}, R.~{Paoletti}, J.~M. {Paredes}, G.~{Pedaletti}, P.~{Pe{\~n}il}, M.~{Peresano}, M.~{Persic}, P.~G. {Prada Moroni}, E.~{Prandini}, I.~{Puljak}, J.~R. {Garcia}, W.~{Rhode}, M.~{Rib{\'o}}, J.~{Rico}, C.~{Righi}, A.~{Rugliancich}, L.~{Saha}, N.~{Sahakyan}, T.~{Saito}, K.~{Satalecka}, T.~{Schweizer}, J.~{Sitarek}, I.~{{\v{S}}nidari{\'c}}, D.~{Sobczynska}, A.~{Somero}, A.~{Stamerra}, M.~{Strzys}, {Periastron
  Observations of TeV Gamma-Ray Emission from a Binary System with a 50-year Period}, ApJL 867~(1) (2018) L19.
\newblock \href {http://arxiv.org/abs/1810.05271} {\path{arXiv:1810.05271}}, \href {https://doi.org/10.3847/2041-8213/aae70e} {\path{doi:10.3847/2041-8213/aae70e}}.

\bibitem{2016ApJ...829..105C}
R.~H.~D. {Corbet}, L.~{Chomiuk}, M.~J. {Coe}, J.~B. {Coley}, G.~{Dubus}, P.~G. {Edwards}, P.~{Martin}, V.~A. {McBride}, J.~{Stevens}, J.~{Strader}, L.~J. {Townsend}, A.~{Udalski}, {A Luminous Gamma-ray Binary in the Large Magellanic Cloud}, ApJ 829~(2) (2016) 105.
\newblock \href {http://arxiv.org/abs/1608.06647} {\path{arXiv:1608.06647}}, \href {https://doi.org/10.3847/0004-637X/829/2/105} {\path{doi:10.3847/0004-637X/829/2/105}}.

\bibitem{2007ApJ...665L..51A}
J.~{Albert}, E.~{Aliu}, H.~{Anderhub}, P.~{Antoranz}, A.~{Armada}, C.~{Baixeras}, J.~A. {Barrio}, H.~{Bartko}, D.~{Bastieri}, J.~K. {Becker}, W.~{Bednarek}, K.~{Berger}, C.~{Bigongiari}, A.~{Biland}, R.~K. {Bock}, P.~{Bordas}, V.~{Bosch-Ramon}, T.~{Bretz}, I.~{Britvitch}, M.~{Camara}, E.~{Carmona}, A.~{Chilingarian}, J.~A. {Coarasa}, S.~{Commichau}, J.~L. {Contreras}, J.~{Cortina}, M.~T. {Costado}, V.~{Curtef}, V.~{Danielyan}, F.~{Dazzi}, A.~{De Angelis}, C.~{Delgado}, R.~{de los Reyes}, B.~{De Lotto}, E.~{Domingo-Santamar{\'\i}a}, D.~{Dorner}, M.~{Doro}, M.~{Errando}, M.~{Fagiolini}, D.~{Ferenc}, E.~{Fern{\'a}ndez}, R.~{Firpo}, J.~{Flix}, M.~V. {Fonseca}, L.~{Font}, M.~{Fuchs}, N.~{Galante}, R.~J. {Garc{\'\i}a-L{\'o}pez}, M.~{Garczarczyk}, M.~{Gaug}, M.~{Giller}, F.~{Goebel}, D.~{Hakobyan}, M.~{Hayashida}, T.~{Hengstebeck}, A.~{Herrero}, D.~{H{\"o}hne}, J.~{Hose}, C.~C. {Hsu}, P.~{Jacon}, T.~{Jogler}, R.~{Kosyra}, D.~{Kranich}, R.~{Kritzer}, A.~{Laille}, E.~{Lindfors}, S.~{Lombardi}, F.~{Longo},
  J.~{L{\'o}pez}, M.~{L{\'o}pez}, E.~{Lorenz}, P.~{Majumdar}, G.~{Maneva}, K.~{Mannheim}, O.~{Mansutti}, M.~{Mariotti}, M.~{Mart{\'\i}nez}, D.~{Mazin}, C.~{Merck}, M.~{Meucci}, M.~{Meyer}, J.~M. {Miranda}, R.~{Mirzoyan}, S.~{Mizobuchi}, A.~{Moralejo}, D.~{Nieto}, K.~{Nilsson}, J.~{Ninkovic}, E.~{O{\~n}a-Wilhelmi}, N.~{Otte}, I.~{Oya}, M.~{Panniello}, R.~{Paoletti}, J.~M. {Paredes}, M.~{Pasanen}, D.~{Pascoli}, F.~{Pauss}, R.~{Pegna}, M.~{Persic}, L.~{Peruzzo}, A.~{Piccioli}, E.~{Prandini}, N.~{Puchades}, A.~{Raymers}, W.~{Rhode}, M.~{Rib{\'o}}, J.~{Rico}, M.~{Rissi}, A.~{Robert}, S.~{R{\"u}gamer}, A.~{Saggion}, T.~{Saito}, A.~{S{\'a}nchez}, P.~{Sartori}, V.~{Scalzotto}, V.~{Scapin}, R.~{Schmitt}, T.~{Schweizer}, M.~{Shayduk}, K.~{Shinozaki}, S.~N. {Shore}, N.~{Sidro}, A.~{Sillanp{\"a}{\"a}}, D.~{Sobczynska}, A.~{Stamerra}, L.~S. {Stark}, L.~{Takalo}, P.~{Temnikov}, D.~{Tescaro}, M.~{Teshima}, D.~F. {Torres}, N.~{Turini}, H.~{Vankov}, V.~{Vitale}, R.~M. {Wagner}, T.~{Wibig}, W.~{Wittek}, F.~{Zandanel},
  R.~{Zanin}, J.~{Zapatero}, {Very High Energy Gamma-Ray Radiation from the Stellar Mass Black Hole Binary Cygnus X-1}, ApJL 665~(1) (2007) L51--L54.
\newblock \href {http://arxiv.org/abs/0706.1505} {\path{arXiv:0706.1505}}, \href {https://doi.org/10.1086/521145} {\path{doi:10.1086/521145}}.

\bibitem{2009Sci...326.1512F}
{Fermi LAT Collaboration}, A.~A. {Abdo}, M.~{Ackermann}, M.~{Ajello}, M.~{Axelsson}, L.~{Baldini}, J.~{Ballet}, G.~{Barbiellini}, D.~{Bastieri}, B.~M. {Baughman}, K.~{Bechtol}, R.~{Bellazzini}, B.~{Berenji}, R.~D. {Blandford}, E.~D. {Bloom}, E.~{Bonamente}, A.~W. {Borgland}, A.~{Brez}, M.~{Brigida}, P.~{Bruel}, T.~H. {Burnett}, S.~{Buson}, G.~A. {Caliandro}, R.~A. {Cameron}, P.~A. {Caraveo}, J.~M. {Casandjian}, C.~{Cecchi}, {\"O}.~{{\c{C}}elik}, S.~{Chaty}, C.~C. {Cheung}, J.~{Chiang}, S.~{Ciprini}, R.~{Claus}, J.~{Cohen-Tanugi}, L.~R. {Cominsky}, J.~{Conrad}, S.~{Corbel}, R.~{Corbet}, C.~D. {Dermer}, F.~{de Palma}, S.~W. {Digel}, E.~{do Couto e Silva}, P.~S. {Drell}, R.~{Dubois}, G.~{Dubus}, D.~{Dumora}, C.~{Farnier}, C.~{Favuzzi}, S.~J. {Fegan}, W.~B. {Focke}, P.~{Fortin}, M.~{Frailis}, P.~{Fusco}, F.~{Gargano}, N.~{Gehrels}, S.~{Germani}, G.~{Giavitto}, B.~{Giebels}, N.~{Giglietto}, F.~{Giordano}, T.~{Glanzman}, G.~{Godfrey}, I.~A. {Grenier}, M.~H. {Grondin}, J.~E. {Grove}, L.~{Guillemot}, S.~{Guiriec},
  Y.~{Hanabata}, A.~K. {Harding}, M.~{Hayashida}, E.~{Hays}, A.~B. {Hill}, L.~{Hjalmarsdotter}, D.~{Horan}, R.~E. {Hughes}, M.~S. {Jackson}, G.~{J{\'o}hannesson}, A.~S. {Johnson}, T.~J. {Johnson}, W.~N. {Johnson}, T.~{Kamae}, H.~{Katagiri}, N.~{Kawai}, M.~{Kerr}, J.~{Kn{\"o}dlseder}, M.~L. {Kocian}, E.~{Koerding}, M.~{Kuss}, J.~{Lande}, J.~{Latronico}, M.~{Lemoine-Goumard}, F.~{Longo}, F.~{Loparco}, B.~{Lott}, M.~N. {Lovellette}, P.~{Lubrano}, G.~M. {Madejski}, A.~{Makeev}, L.~{Marchand}, M.~{Marelli}, W.~{Max-Moerbeck}, M.~N. {Mazziotta}, N.~{McColl}, J.~E. {McEnery}, C.~{Meurer}, P.~F. {Michelson}, S.~{Migliari}, W.~{Mitthumsiri}, T.~{Mizuno}, C.~{Monte}, M.~E. {Monzani}, A.~{Morselli}, I.~V. {Moskalenko}, S.~{Murgia}, P.~L. {Nolan}, J.~P. {Norris}, E.~{Nuss}, T.~{Ohsugi}, N.~{Omodei}, R.~A. {Ong}, J.~F. {Ormes}, D.~{Paneque}, D.~{Parent}, V.~{Pelassa}, M.~{Pepe}, M.~{Pesce-Rollins}, F.~{Piron}, G.~{Pooley}, T.~A. {Porter}, K.~{Pottschmidt}, S.~{Rain{\`o}}, R.~{Rando}, P.~S. {Ray}, M.~{Razzano}, N.~{Rea},
  A.~{Readhead}, A.~{Reimer}, O.~{Reimer}, J.~L. {Richards}, L.~S. {Rochester}, J.~{Rodriguez}, A.~Y. {Rodriguez}, R.~W. {Romani}, F.~{Ryde}, H.~F.~W. {Sadrozinski}, A.~{Sander}, P.~M. {Saz Parkinson}, C.~{Sgr{\`o}}, E.~J. {Siskind}, D.~A. {Smith}, P.~D. {Smith}, P.~{Spinelli}, J.~L. {Starck}, M.~{Stevenson}, M.~S. {Strickman}, D.~J. {Suson}, H.~{Takahashi}, T.~{Tanaka}, J.~B. {Thayer}, D.~J. {Thompson}, L.~{Tibaldo}, J.~A. {Tomsick}, D.~F. {Torres}, G.~{Tosti}, A.~{Tramacere}, Y.~{Uchiyama}, T.~L. {Usher}, V.~{Vasileiou}, N.~{Vilchez}, V.~{Vitale}, A.~P. {Waite}, P.~{Wang}, J.~{Wilms}, B.~L. {Winer}, K.~W. {Wood}, T.~{Ylinen}, M.~{Ziegler}, {Modulated High-Energy Gamma-Ray Emission from the Microquasar Cygnus X-3}, Science 326~(5959) (2009) 1512.
\newblock \href {https://doi.org/10.1126/science.1182174} {\path{doi:10.1126/science.1182174}}.

\bibitem{2007A&A...469L...1A}
F.~A. {Aharonian}, A.~G. {Akhperjanian}, A.~R. {Bazer-Bachi}, B.~{Behera}, M.~{Beilicke}, W.~{Benbow}, D.~{Berge}, K.~{Bernl{\"o}hr}, C.~{Boisson}, O.~{Bolz}, V.~{Borrel}, I.~{Braun}, E.~{Brion}, A.~M. {Brown}, R.~{B{\"u}hler}, I.~{B{\"u}sching}, T.~{Boutelier}, S.~{Carrigan}, P.~M. {Chadwick}, L.~M. {Chounet}, G.~{Coignet}, R.~{Cornils}, L.~{Costamante}, B.~{Degrange}, H.~J. {Dickinson}, A.~{Djannati-Ata{\"\i}}, W.~{Domainko}, L.~{O'C. Drury}, G.~{Dubus}, K.~{Egberts}, D.~{Emmanoulopoulos}, P.~{Espigat}, C.~{Farnier}, F.~{Feinstein}, A.~{Fiasson}, A.~{F{\"o}rster}, G.~{Fontaine}, S.~{Funk}, S.~{Funk}, M.~{F{\"u}{\ss}ling}, Y.~A. {Gallant}, B.~{Giebels}, J.~F. {Glicenstein}, B.~{Gl{\"u}ck}, P.~{Goret}, C.~{Hadjichristidis}, D.~{Hauser}, M.~{Hauser}, G.~{Heinzelmann}, G.~{Henri}, G.~{Hermann}, J.~A. {Hinton}, A.~{Hoffmann}, W.~{Hofmann}, M.~{Holleran}, S.~{Hoppe}, D.~{Horns}, A.~{Jacholkowska}, O.~C. {de Jager}, E.~{Kendziorra}, M.~{Kerschhaggl}, B.~{Kh{\'e}lifi}, N.~{Komin}, K.~{Kosack}, G.~{Lamanna}, I.~J.
  {Latham}, R.~{Le Gallou}, A.~{Lemi{\`e}re}, M.~{Lemoine-Goumard}, T.~{Lohse}, J.~M. {Martin}, O.~{Martineau-Huynh}, A.~{Marcowith}, C.~{Masterson}, G.~{Maurin}, T.~J.~L. {McComb}, E.~{Moulin}, M.~{de Naurois}, D.~{Nedbal}, S.~J. {Nolan}, A.~{Noutsos}, J.~P. {Olive}, K.~J. {Orford}, J.~L. {Osborne}, M.~{Panter}, G.~{Pedaletti}, G.~{Pelletier}, P.~O. {Petrucci}, S.~{Pita}, G.~{P{\"u}hlhofer}, M.~{Punch}, S.~{Ranchon}, B.~C. {Raubenheimer}, M.~{Raue}, S.~M. {Rayner}, O.~{Reimer}, J.~{Ripken}, L.~{Rob}, L.~{Rolland}, S.~{Rosier-Lees}, G.~{Rowell}, J.~{Ruppel}, V.~{Sahakian}, A.~{Santangelo}, L.~{Saug{\'e}}, S.~{Schlenker}, R.~{Schlickeiser}, R.~{Schr{\"o}der}, U.~{Schwanke}, S.~{Schwarzburg}, S.~{Schwemmer}, A.~{Shalchi}, H.~{Sol}, D.~{Spangler}, R.~{Steenkamp}, C.~{Stegmann}, G.~{Superina}, P.~H. {Tam}, J.~P. {Tavernet}, R.~{Terrier}, M.~{Tluczykont}, C.~{van Eldik}, G.~{Vasileiadis}, C.~{Venter}, J.~P. {Vialle}, P.~{Vincent}, H.~J. {V{\"o}lk}, S.~J. {Wagner}, M.~{Ward}, Y.~{Moriguchi}, Y.~{Fukui}, {Discovery
  of a point-like very-high-energy {\ensuremath{\gamma}}-ray source in Monoceros}, A\&A 469~(1) (2007) L1--L4.
\newblock \href {http://arxiv.org/abs/0704.0171} {\path{arXiv:0704.0171}}, \href {https://doi.org/10.1051/0004-6361:20077299} {\path{doi:10.1051/0004-6361:20077299}}.

\bibitem{2009ApJ...690L.101H}
J.~A. {Hinton}, J.~L. {Skilton}, S.~{Funk}, J.~{Brucker}, F.~A. {Aharonian}, G.~{Dubus}, A.~{Fiasson}, Y.~{Gallant}, W.~{Hofmann}, A.~{Marcowith}, O.~{Reimer}, {HESS J0632+057: A New Gamma-Ray Binary?}, ApJL 690~(2) (2009) L101--L104.
\newblock \href {http://arxiv.org/abs/0809.0584} {\path{arXiv:0809.0584}}, \href {https://doi.org/10.1088/0004-637X/690/2/L101} {\path{doi:10.1088/0004-637X/690/2/L101}}.

\bibitem{2010ApJ...724..306A}
C.~{Aragona}, M.~V. {McSwain}, M.~{De Becker}, {HD 259440: The Proposed Optical Counterpart of the {\ensuremath{\gamma}}-ray Binary HESS J0632+057}, ApJ 724~(1) (2010) 306--312.
\newblock \href {http://arxiv.org/abs/1009.2100} {\path{arXiv:1009.2100}}, \href {https://doi.org/10.1088/0004-637X/724/1/306} {\path{doi:10.1088/0004-637X/724/1/306}}.

\bibitem{2009ApJ...698L..94A}
V.~A. {Acciari}, E.~{Aliu}, T.~{Arlen}, M.~{Beilicke}, W.~{Benbow}, D.~{Boltuch}, S.~M. {Bradbury}, J.~H. {Buckley}, V.~{Bugaev}, K.~{Byrum}, A.~{Cannon}, A.~{Cesarini}, A.~{Cesarini}, Y.~C. {Chow}, L.~{Ciupik}, P.~{Cogan}, R.~{Dickherber}, C.~{Duke}, T.~{Ergin}, A.~{Falcone}, S.~J. {Fegan}, J.~P. {Finley}, G.~{Finnegan}, P.~{Fortin}, L.~{Fortson}, A.~{Furniss}, K.~{Gibbs}, G.~H. {Gillanders}, J.~{Grube}, R.~{Guenette}, G.~{Gyuk}, D.~{Hanna}, J.~{Holder}, D.~{Horan}, C.~M. {Hui}, T.~B. {Humensky}, A.~{Imran}, P.~{Kaaret}, N.~{Karlsson}, M.~{Kertzman}, D.~{Kieda}, J.~{Kildea}, A.~{Konopelko}, H.~{Krawczynski}, F.~{Krennrich}, M.~J. {Lang}, S.~{LeBohec}, S.~{LeBohec}, G.~{Maier}, A.~{McCann}, M.~{McCutcheon}, J.~{Millis}, J.~{Millis}, P.~{Moriarty}, R.~{Mukherjee}, R.~A. {Ong}, A.~N. {Otte}, D.~{Pandel}, J.~S. {Perkins}, D.~{Petry}, M.~{Pohl}, J.~{Quinn}, K.~{Ragan}, L.~C. {Reyes}, P.~T. {Reynolds}, H.~J. {Rose}, M.~{Schroedter}, G.~H. {Sembroski}, A.~W. {Smith}, D.~{Steele}, S.~{Swordy}, M.~{Theiling}, J.~A.
  {Toner}, A.~{Varlotta}, V.~V. {Vassiliev}, S.~{Vincent}, R.~G. {Wagner}, S.~P. {Wakely}, J.~E. {Ward}, T.~C. {Weekes}, A.~{Weinstein}, T.~{Weisgarber}, D.~A. {Williams}, S.~{Wissel}, M.~{Wood}, {Evidence for Long-Term Gamma-Ray and X-Ray Variability from the Unidentified TeV Source HESS J0632+057}, ApJL 698~(2) (2009) L94--L97.
\newblock \href {http://arxiv.org/abs/0905.3139} {\path{arXiv:0905.3139}}, \href {https://doi.org/10.1088/0004-637X/698/2/L94} {\path{doi:10.1088/0004-637X/698/2/L94}}.

\bibitem{2000IAUC.7432....3V}
W.~{Voges}, B.~{Aschenbach}, T.~{Boller}, H.~{Brauninger}, U.~{Briel}, W.~{Burkert}, K.~{Dennerl}, J.~{Englhauser}, R.~{Gruber}, F.~{Haberl}, G.~{Hartner}, G.~{Hasinger}, E.~{Pfeffermann}, W.~{Pietsch}, P.~{Predehl}, J.~{Schmitt}, J.~{Trumper}, U.~{Zimmermann}, {Rosat All-Sky Survey Faint Source Catalogue}, IAUC 7432 (2000) 3.

\bibitem{2017ApJ...846..169L}
J.~{Li}, D.~F. {Torres}, K.~S. {Cheng}, E.~{de O{\~n}a Wilhelmi}, P.~{Kretschmar}, X.~{Hou}, J.~{Takata}, {GeV Detection of HESS J0632+057}, ApJ 846~(2) (2017) 169.
\newblock \href {http://arxiv.org/abs/1707.04280} {\path{arXiv:1707.04280}}, \href {https://doi.org/10.3847/1538-4357/aa7ff7} {\path{doi:10.3847/1538-4357/aa7ff7}}.

\bibitem{2020ApJS..247...33A}
S.~{Abdollahi}, F.~{Acero}, M.~{Ackermann}, M.~{Ajello}, W.~B. {Atwood}, M.~{Axelsson}, L.~{Baldini}, J.~{Ballet}, G.~{Barbiellini}, D.~{Bastieri}, J.~{Becerra Gonzalez}, R.~{Bellazzini}, A.~{Berretta}, E.~{Bissaldi}, R.~D. {Blandford}, E.~D. {Bloom}, R.~{Bonino}, E.~{Bottacini}, T.~J. {Brandt}, J.~{Bregeon}, P.~{Bruel}, R.~{Buehler}, T.~H. {Burnett}, S.~{Buson}, R.~A. {Cameron}, R.~{Caputo}, P.~A. {Caraveo}, J.~M. {Casandjian}, D.~{Castro}, E.~{Cavazzuti}, E.~{Charles}, S.~{Chaty}, S.~{Chen}, C.~C. {Cheung}, G.~{Chiaro}, S.~{Ciprini}, J.~{Cohen-Tanugi}, L.~R. {Cominsky}, J.~{Coronado-Bl{\'a}zquez}, D.~{Costantin}, A.~{Cuoco}, S.~{Cutini}, F.~{D'Ammando}, M.~{DeKlotz}, P.~{de la Torre Luque}, F.~{de Palma}, A.~{Desai}, S.~W. {Digel}, N.~{Di Lalla}, M.~{Di Mauro}, L.~{Di Venere}, A.~{Dom{\'\i}nguez}, D.~{Dumora}, F.~{Fana Dirirsa}, S.~J. {Fegan}, E.~C. {Ferrara}, A.~{Franckowiak}, Y.~{Fukazawa}, S.~{Funk}, P.~{Fusco}, F.~{Gargano}, D.~{Gasparrini}, N.~{Giglietto}, P.~{Giommi}, F.~{Giordano}, M.~{Giroletti},
  T.~{Glanzman}, D.~{Green}, I.~A. {Grenier}, S.~{Griffin}, M.~H. {Grondin}, J.~E. {Grove}, S.~{Guiriec}, A.~K. {Harding}, K.~{Hayashi}, E.~{Hays}, J.~W. {Hewitt}, D.~{Horan}, G.~{J{\'o}hannesson}, T.~J. {Johnson}, T.~{Kamae}, M.~{Kerr}, D.~{Kocevski}, M.~{Kovac'evic'}, M.~{Kuss}, D.~{Landriu}, S.~{Larsson}, L.~{Latronico}, M.~{Lemoine-Goumard}, J.~{Li}, I.~{Liodakis}, F.~{Longo}, F.~{Loparco}, B.~{Lott}, M.~N. {Lovellette}, P.~{Lubrano}, G.~M. {Madejski}, S.~{Maldera}, D.~{Malyshev}, A.~{Manfreda}, E.~J. {Marchesini}, L.~{Marcotulli}, G.~{Mart{\'\i}-Devesa}, P.~{Martin}, F.~{Massaro}, M.~N. {Mazziotta}, J.~E. {McEnery}, I.~{Mereu}, M.~{Meyer}, P.~F. {Michelson}, N.~{Mirabal}, T.~{Mizuno}, M.~E. {Monzani}, A.~{Morselli}, I.~V. {Moskalenko}, M.~{Negro}, E.~{Nuss}, R.~{Ojha}, N.~{Omodei}, M.~{Orienti}, E.~{Orlando}, J.~F. {Ormes}, M.~{Palatiello}, V.~S. {Paliya}, D.~{Paneque}, Z.~{Pei}, H.~{Pe{\~n}a-Herazo}, J.~S. {Perkins}, M.~{Persic}, M.~{Pesce-Rollins}, V.~{Petrosian}, L.~{Petrov}, F.~{Piron}, H.~{Poon},
  T.~A. {Porter}, G.~{Principe}, S.~{Rain{\`o}}, R.~{Rando}, M.~{Razzano}, S.~{Razzaque}, A.~{Reimer}, O.~{Reimer}, Q.~{Remy}, T.~{Reposeur}, R.~W. {Romani}, P.~M. {Saz Parkinson}, F.~K. {Schinzel}, D.~{Serini}, C.~{Sgr{\`o}}, E.~J. {Siskind}, D.~A. {Smith}, G.~{Spandre}, P.~{Spinelli}, A.~W. {Strong}, D.~J. {Suson}, H.~{Tajima}, M.~N. {Takahashi}, D.~{Tak}, J.~B. {Thayer}, D.~J. {Thompson}, L.~{Tibaldo}, D.~F. {Torres}, E.~{Torresi}, J.~{Valverde}, B.~{Van Klaveren}, P.~{van Zyl}, K.~{Wood}, M.~{Yassine}, G.~{Zaharijas}, {Fermi Large Area Telescope Fourth Source Catalog}, ApJS 247~(1) (2020) 33.
\newblock \href {http://arxiv.org/abs/1902.10045} {\path{arXiv:1902.10045}}, \href {https://doi.org/10.3847/1538-4365/ab6bcb} {\path{doi:10.3847/1538-4365/ab6bcb}}.

\bibitem{2010ApJ...708L..52F}
A.~D. {Falcone}, J.~{Grube}, J.~{Hinton}, J.~{Holder}, G.~{Maier}, R.~{Mukherjee}, J.~{Skilton}, M.~{Stroh}, {Probing the Nature of the Unidentified TeV Gamma-Ray Source Hess J0632+057 with Swift}, ApJL 708~(1) (2010) L52--L56.
\newblock \href {http://arxiv.org/abs/0912.0941} {\path{arXiv:0912.0941}}, \href {https://doi.org/10.1088/2041-8205/708/1/L52} {\path{doi:10.1088/2041-8205/708/1/L52}}.

\bibitem{2011ApJ...737L..11B}
S.~D. {Bongiorno}, A.~D. {Falcone}, M.~{Stroh}, J.~{Holder}, J.~L. {Skilton}, J.~A. {Hinton}, N.~{Gehrels}, J.~{Grube}, {A New TeV Binary: The Discovery of an Orbital Period in HESS J0632+057}, ApJL 737~(1) (2011) L11.
\newblock \href {http://arxiv.org/abs/1104.4519} {\path{arXiv:1104.4519}}, \href {https://doi.org/10.1088/2041-8205/737/1/L11} {\path{doi:10.1088/2041-8205/737/1/L11}}.

\bibitem{2014ApJ...780..168A}
E.~{Aliu}, S.~{Archambault}, T.~{Aune}, B.~{Behera}, M.~{Beilicke}, W.~{Benbow}, K.~{Berger}, R.~{Bird}, A.~{Bouvier}, J.~H. {Buckley}, V.~{Bugaev}, K.~{Byrum}, M.~{Cerruti}, X.~{Chen}, L.~{Ciupik}, M.~P. {Connolly}, W.~{Cui}, C.~{Duke}, J.~{Dumm}, M.~{Errando}, A.~{Falcone}, S.~{Federici}, Q.~{Feng}, J.~P. {Finley}, P.~{Fortin}, L.~{Fortson}, A.~{Furniss}, N.~{Galante}, G.~H. {Gillanders}, S.~{Griffin}, S.~T. {Griffiths}, J.~{Grube}, G.~{Gyuk}, D.~{Hanna}, J.~{Holder}, G.~{Hughes}, T.~B. {Humensky}, P.~{Kaaret}, M.~{Kertzman}, Y.~{Khassen}, D.~{Kieda}, H.~{Krawczynski}, F.~{Krennrich}, M.~J. {Lang}, A.~S. {Madhavan}, G.~{Maier}, P.~{Majumdar}, A.~{McCann}, P.~{Moriarty}, R.~{Mukherjee}, D.~{Nieto}, A.~{O'Faol{\'a}in de Bhr{\'o}ithe}, R.~A. {Ong}, A.~N. {Otte}, N.~{Park}, J.~S. {Perkins}, M.~{Pohl}, A.~{Popkow}, H.~{Prokoph}, J.~{Quinn}, K.~{Ragan}, J.~{Rajotte}, L.~C. {Reyes}, P.~T. {Reynolds}, G.~T. {Richards}, E.~{Roache}, J.~{Rousselle}, G.~H. {Sembroski}, F.~{Sheidaei}, C.~{Skole}, A.~W. {Smith},
  D.~{Staszak}, M.~{Stroh}, I.~{Telezhinsky}, M.~{Theiling}, J.~V. {Tucci}, J.~{Tyler}, A.~{Varlotta}, S.~{Vincent}, S.~P. {Wakely}, A.~{Weinstein}, R.~{Welsing}, D.~A. {Williams}, A.~{Zajczyk}, B.~{Zitzer}, {VERITAS Collaboration}, A.~{Abramowski}, F.~{Aharonian}, F.~A. {Benkhali}, A.~G. {Akhperjanian}, E.~{Ang{\"u}ner}, G.~{Anton}, S.~{Balenderan}, A.~{Balzer}, A.~{Barnacka}, Y.~{Becherini}, J.~{Becker Tjus}, K.~{Bernl{\"o}hr}, E.~{Birsin}, E.~{Bissaldi}, J.~{Biteau}, M.~{B{\"o}ttcher}, C.~{Boisson}, J.~{Bolmont}, P.~{Bordas}, J.~{Brucker}, F.~{Brun}, P.~{Brun}, T.~{Bulik}, S.~{Carrigan}, S.~{Casanova}, M.~{Cerruti}, P.~M. {Chadwick}, R.~{Chalme-Calvet}, R.~C.~G. {Chaves}, A.~{Cheesebrough}, M.~{Chr{\'e}tien}, S.~{Colafrancesco}, G.~{Cologna}, J.~{Conrad}, C.~{Couturier}, M.~{Dalton}, M.~K. {Daniel}, I.~D. {Davids}, B.~{Degrange}, C.~{Deil}, P.~{deWilt}, H.~J. {Dickinson}, A.~{Djannati-Ata{\"\i}}, W.~{Domainko}, L.~{O'C. Drury}, G.~{Dubus}, K.~{Dutson}, J.~{Dyks}, M.~{Dyrda}, T.~{Edwards}, K.~{Egberts},
  P.~{Eger}, P.~{Espigat}, C.~{Farnier}, S.~{Fegan}, F.~{Feinstein}, M.~V. {Fernandes}, D.~{Fernandez}, A.~{Fiasson}, G.~{Fontaine}, A.~{F{\"o}rster}, M.~{F{\"u}ssling}, M.~{Gajdus}, Y.~A. {Gallant}, T.~{Garrigoux}, G.~{Giavitto}, B.~{Giebels}, J.~F. {Glicenstein}, M.~H. {Grondin}, M.~{Grudzi{\'n}ska}, S.~{H{\"a}ffner}, J.~{Hahn}, J.~{Harris}, G.~{Heinzelmann}, G.~{Henri}, G.~{Hermann}, O.~{Hervet}, A.~{Hillert}, J.~A. {Hinton}, W.~{Hofmann}, P.~{Hofverberg}, M.~{Holler}, D.~{Horns}, A.~{Jacholkowska}, C.~{Jahn}, M.~{Jamrozy}, M.~{Janiak}, F.~{Jankowsky}, I.~{Jung}, M.~A. {Kastendieck}, K.~{Katarzy{\'n}ski}, U.~{Katz}, S.~{Kaufmann}, B.~{Kh{\'e}lifi}, M.~{Kieffer}, S.~{Klepser}, D.~{Klochkov}, W.~{Klu{\'z}niak}, T.~{Kneiske}, D.~{Kolitzus}, N.~{Komin}, K.~{Kosack}, S.~{Krakau}, F.~{Krayzel}, P.~P. {Kr{\"u}ger}, H.~{Laffon}, G.~{Lamanna}, J.~{Lefaucheur}, A.~{Lemi{\`e}re}, M.~{Lemoine-Goumard}, J.~P. {Lenain}, D.~{Lennarz}, T.~{Lohse}, A.~{Lopatin}, C.~C. {Lu}, V.~{Marandon}, A.~{Marcowith}, R.~{Marx},
  G.~{Maurin}, N.~{Maxted}, M.~{Mayer}, T.~J.~L. {McComb}, J.~{M{\'e}hault}, U.~{Menzler}, M.~{Meyer}, R.~{Moderski}, M.~{Mohamed}, E.~{Moulin}, T.~{Murach}, C.~L. {Naumann}, M.~{de Naurois}, J.~{Niemiec}, S.~J. {Nolan}, L.~{Oakes}, S.~{Ohm}, E.~{de O{\~n}a Wilhelmi}, B.~{Opitz}, M.~{Ostrowski}, I.~{Oya}, M.~{Panter}, R.~D. {Parsons}, M.~{Paz Arribas}, N.~W. {Pekeur}, G.~{Pelletier}, J.~{Perez}, P.~O. {Petrucci}, B.~{Peyaud}, S.~{Pita}, H.~{Poon}, G.~{P{\"u}hlhofer}, M.~{Punch}, A.~{Quirrenbach}, S.~{Raab}, M.~{Raue}, A.~{Reimer}, O.~{Reimer}, M.~{Renaud}, R.~{de los Reyes}, F.~{Rieger}, L.~{Rob}, C.~{Romoli}, S.~{Rosier-Lees}, G.~{Rowell}, B.~{Rudak}, C.~B. {Rulten}, V.~{Sahakian}, D.~A. {Sanchez}, A.~{Santangelo}, R.~{Schlickeiser}, F.~{Sch{\"u}ssler}, A.~{Schulz}, U.~{Schwanke}, S.~{Schwarzburg}, S.~{Schwemmer}, H.~{Sol}, G.~{Spengler}, F.~{Spies}, {\L}.~{Stawarz}, R.~{Steenkamp}, C.~{Stegmann}, F.~{Stinzing}, K.~{Stycz}, I.~{Sushch}, A.~{Szostek}, J.~P. {Tavernet}, T.~{Tavernier}, A.~M. {Taylor},
  R.~{Terrier}, M.~{Tluczykont}, C.~{Trichard}, K.~{Valerius}, C.~{van Eldik}, G.~{Vasileiadis}, C.~{Venter}, A.~{Viana}, P.~{Vincent}, H.~J. {V{\"o}lk}, F.~{Volpe}, M.~{Vorster}, S.~J. {Wagner}, P.~{Wagner}, M.~{Ward}, M.~{Weidinger}, Q.~{Weitzel}, R.~{White}, A.~{Wierzcholska}, P.~{Willmann}, A.~{W{\"o}rnlein}, D.~{Wouters}, M.~{Zacharias}, A.~{Zajczyk}, A.~A. {Zdziarski}, A.~{Zech}, H.~S. {Zechlin}, {H.~E.~S.~S. Collaboration}, {Long-term TeV and X-Ray Observations of the Gamma-Ray Binary HESS J0632+057}, ApJ 780~(2) (2014) 168.
\newblock \href {http://arxiv.org/abs/1311.6083} {\path{arXiv:1311.6083}}, \href {https://doi.org/10.1088/0004-637X/780/2/168} {\path{doi:10.1088/0004-637X/780/2/168}}.

\bibitem{2021ApJ...923..241A}
C.~B. {Adams}, W.~{Benbow}, A.~{Brill}, J.~H. {Buckley}, M.~{Capasso}, A.~J. {Chromey}, M.~{Errando}, A.~{Falcone}, K.~{A Farrell}, Q.~{Feng}, J.~P. {Finley}, G.~{M Foote}, L.~{Fortson}, A.~{Furniss}, A.~{Gent}, G.~H. {Gillanders}, C.~{Giuri}, O.~{Gueta}, D.~{Hanna}, T.~{Hassan}, O.~{Hervet}, J.~{Holder}, B.~{Hona}, T.~B. {Humensky}, W.~{Jin}, P.~{Kaaret}, M.~{Kertzman}, D.~{Kieda}, T.~{K Kleiner}, F.~{Krennrich}, S.~{Kumar}, M.~J. {Lang}, M.~{Lundy}, G.~{Maier}, C.~{E McGrath}, P.~{Moriarty}, R.~{Mukherjee}, D.~{Nieto}, M.~{Nievas-Rosillo}, S.~{O'Brien}, R.~A. {Ong}, A.~N. {Otte}, N.~{Park}, S.~{Patel}, K.~{Pfrang}, A.~{Pichel}, M.~{Pohl}, R.~R. {Prado}, J.~{Quinn}, K.~{Ragan}, P.~T. {Reynolds}, D.~{Ribeiro}, E.~{Roache}, A.~C. {Rovero}, J.~L. {Ryan}, M.~{Santander}, S.~{Schlenstedt}, G.~H. {Sembroski}, R.~{Shang}, D.~{Tak}, V.~V. {Vassiliev}, A.~{Weinstein}, D.~A. {Williams}, T.~{J Williamson}, T.~{J Williamson}, V.~A. {Acciari}, S.~{Ansoldi}, L.~A. {Antonelli}, A.~{Arbet Engels}, M.~{Artero}, K.~{Asano},
  D.~{Baack}, A.~{Babi{\'c}}, A.~{Baquero}, U.~{Barres de Almeida}, J.~A. {Barrio}, I.~{Batkovi{\'c}}, J.~{Becerra Gonz{\'a}lez}, W.~{Bednarek}, L.~{Bellizzi}, E.~{Bernardini}, M.~{Bernardos}, A.~{Berti}, J.~{Besenrieder}, W.~{Bhattacharyya}, C.~{Bigongiari}, A.~{Biland}, O.~{Blanch}, H.~{B{\"o}kenkamp}, G.~{Bonnoli}, {\v{Z}}.~{Bo{\v{s}}njak}, G.~{Busetto}, R.~{Carosi}, G.~{Ceribella}, M.~{Cerruti}, Y.~{Chai}, A.~{Chilingarian}, S.~{Cikota}, S.~M. {Colak}, E.~{Colombo}, J.~L. {Contreras}, J.~{Cortina}, S.~{Covino}, G.~{D'Amico}, V.~{D'Elia}, P.~{Da Vela}, F.~{Dazzi}, A.~{De Angelis}, B.~{De Lotto}, M.~{Delfino}, J.~{Delgado}, C.~{Delgado Mendez}, D.~{Depaoli}, F.~{Di Pierro}, L.~{Di Venere}, E.~{Do Souto Espi{\~n}eira}, D.~{Dominis Prester}, A.~{Donini}, D.~{Dorner}, M.~{Doro}, D.~{Elsaesser}, V.~{Fallah Ramazani}, A.~{Fattorini}, M.~V. {Fonseca}, L.~{Font}, C.~{Fruck}, S.~{Fukami}, Y.~{Fukazawa}, R.~J. {Garc{\'\i}a L{\'o}pez}, M.~{Garczarczyk}, S.~{Gasparyan}, M.~{Gaug}, N.~{Giglietto}, F.~{Giordano},
  P.~{Gliwny}, N.~{Godinovi{\'c}}, J.~G. {Green}, D.~{Green}, D.~{Hadasch}, A.~{Hahn}, L.~{Heckmann}, J.~{Herrera}, J.~{Hoang}, D.~{Hrupec}, M.~{H{\"u}tten}, T.~{Inada}, K.~{Ishio}, Y.~{Iwamura}, I.~{Jim{\'e}nez Mart{\'\i}nez}, J.~{Jormanainen}, L.~{Jouvin}, M.~{Karjalainen}, D.~{Kerszberg}, Y.~{Kobayashi}, H.~{Kubo}, J.~{Kushida}, A.~{Lamastra}, D.~{Lelas}, F.~{Leone}, E.~{Lindfors}, L.~{Linhoff}, S.~{Lombardi}, F.~{Longo}, R.~{L{\'o}pez-Coto}, M.~{L{\'o}pez-Moya}, A.~{L{\'o}pez-Oramas}, S.~{Loporchio}, B.~{Machado de Oliveira Fraga}, C.~{Maggio}, P.~{Majumdar}, M.~{Makariev}, M.~{Mallamaci}, G.~{Maneva}, M.~{Manganaro}, K.~{Mannheim}, L.~{Maraschi}, M.~{Mariotti}, M.~{Mart{\'\i}nez}, D.~{Mazin}, S.~{Menchiari}, S.~{Mender}, S.~{Mi{\'c}anovi{\'c}}, D.~{Miceli}, T.~{Miener}, J.~M. {Miranda}, R.~{Mirzoyan}, E.~{Molina}, A.~{Moralejo}, D.~{Morcuende}, V.~{Moreno}, E.~{Moretti}, T.~{Nakamori}, L.~{Nava}, V.~{Neustroev}, C.~{Nigro}, K.~{Nilsson}, K.~{Nishijima}, K.~{Noda}, S.~{Nozaki}, Y.~{Ohtani}, {Observation
  of the Gamma-Ray Binary HESS J0632+057 with the H.E.S.S., MAGIC, and VERITAS Telescopes}, ApJ 923~(2) (2021) 241.
\newblock \href {http://arxiv.org/abs/2109.11894} {\path{arXiv:2109.11894}}, \href {https://doi.org/10.3847/1538-4357/ac29b7} {\path{doi:10.3847/1538-4357/ac29b7}}.

\bibitem{2012MNRAS.421.1103C}
J.~{Casares}, M.~{Rib{\'o}}, I.~{Ribas}, J.~M. {Paredes}, F.~{Vilardell}, I.~{Negueruela}, {On the binary nature of the {\ensuremath{\gamma}}-ray sources AGL J2241+4454 (= MWC 656) and HESS J0632+057 (= MWC 148)}, MNRAS 421~(2) (2012) 1103--1112.
\newblock \href {http://arxiv.org/abs/1201.1726} {\path{arXiv:1201.1726}}, \href {https://doi.org/10.1111/j.1365-2966.2011.20368.x} {\path{doi:10.1111/j.1365-2966.2011.20368.x}}.

\bibitem{2018PASJ...70...61M}
Y.~{Moritani}, T.~{Kawano}, S.~{Chimasu}, A.~{Kawachi}, H.~{Takahashi}, J.~{Takata}, A.~C. {Carciofi}, {Orbital solution leading to an acceptable interpretation for the enigmatic gamma-ray binary HESS J0632+057}, PASJ 70~(4) (2018) 61.
\newblock \href {http://arxiv.org/abs/1804.03831} {\path{arXiv:1804.03831}}, \href {https://doi.org/10.1093/pasj/psy053} {\path{doi:10.1093/pasj/psy053}}.

\bibitem{2025MNRAS.536..166M}
N.~{Matchett}, B.~{van Soelen}, {New insight into the orbital parameters of the gamma-ray binary HESS J0632 + 057}, MNRAS 536~(1) (2025) 166--173.
\newblock \href {http://arxiv.org/abs/2411.12499} {\path{arXiv:2411.12499}}, \href {https://doi.org/10.1093/mnras/stae2597} {\path{doi:10.1093/mnras/stae2597}}.

\bibitem{2020mbhe.confE..45C}
M.~{Chernyakova}, D.~{Malyshev}, {Gamma-ray binaries}, in: Multifrequency Behaviour of High Energy Cosmic Sources - XIII. 3-8 June 2019. Palermo, 2020, p.~45.
\newblock \href {http://arxiv.org/abs/2006.03615} {\path{arXiv:2006.03615}}, \href {https://doi.org/10.22323/1.362.0045} {\path{doi:10.22323/1.362.0045}}.

\bibitem{2015MNRAS.454.1358C}
M.~{Chernyakova}, A.~{Neronov}, B.~{van Soelen}, P.~{Callanan}, L.~{O'Shaughnessy}, I.~{Babyk}, S.~{Tsygankov}, I.~{Vovk}, R.~{Krivonos}, J.~A. {Tomsick}, D.~{Malyshev}, J.~{Li}, K.~{Wood}, D.~{Torres}, S.~{Zhang}, P.~{Kretschmar}, M.~V. {McSwain}, D.~A.~H. {Buckley}, C.~{Koen}, {Multi-wavelength observations of the binary system PSR B1259-63/LS 2883 around the 2014 periastron passage}, MNRAS 454~(2) (2015) 1358--1370.
\newblock \href {http://arxiv.org/abs/1508.01339} {\path{arXiv:1508.01339}}, \href {https://doi.org/10.1093/mnras/stv1988} {\path{doi:10.1093/mnras/stv1988}}.

\bibitem{2019A&A...627A..87C}
A.~M. {Chen}, J.~{Takata}, S.~X. {Yi}, Y.~W. {Yu}, K.~S. {Cheng}, {Modelling multiwavelength emissions from PSR B1259-63/LS 2883: Effects of the stellar disc on shock radiations}, A\&A 627 (2019) A87.
\newblock \href {http://arxiv.org/abs/1904.07527} {\path{arXiv:1904.07527}}, \href {https://doi.org/10.1051/0004-6361/201935166} {\path{doi:10.1051/0004-6361/201935166}}.

\bibitem{2017ApJ...836..241T}
J.~{Takata}, P.~H.~T. {Tam}, C.~W. {Ng}, K.~L. {Li}, A.~K.~H. {Kong}, C.~Y. {Hui}, K.~S. {Cheng}, {High-energy Emissions from the Pulsar/Be Binary System PSR J2032+4127/MT91 213}, ApJ 836~(2) (2017) 241.
\newblock \href {http://arxiv.org/abs/1702.04446} {\path{arXiv:1702.04446}}, \href {https://doi.org/10.3847/1538-4357/aa5c80} {\path{doi:10.3847/1538-4357/aa5c80}}.

\bibitem{2022A&A...658A.153C}
A.~M. {Chen}, J.~{Takata}, {Modelling the correlated keV/TeV light curves of Be/gamma-ray binaries}, A\&A 658 (2022) A153.
\newblock \href {http://arxiv.org/abs/2112.00345} {\path{arXiv:2112.00345}}, \href {https://doi.org/10.1051/0004-6361/202142258} {\path{doi:10.1051/0004-6361/202142258}}.

\bibitem{2023arXiv230712546B}
J.~{Ballet}, P.~{Bruel}, T.~H. {Burnett}, B.~{Lott}, {The Fermi-LAT collaboration}, {Fermi Large Area Telescope Fourth Source Catalog Data Release 4 (4FGL-DR4)}, arXiv e-prints (2023) arXiv:2307.12546\href {http://arxiv.org/abs/2307.12546} {\path{arXiv:2307.12546}}, \href {https://doi.org/10.48550/arXiv.2307.12546} {\path{doi:10.48550/arXiv.2307.12546}}.

\bibitem{1996ApJ...461..396M}
J.~R. {Mattox}, D.~L. {Bertsch}, J.~{Chiang}, B.~L. {Dingus}, S.~W. {Digel}, J.~A. {Esposito}, J.~M. {Fierro}, R.~C. {Hartman}, S.~D. {Hunter}, G.~{Kanbach}, D.~A. {Kniffen}, Y.~C. {Lin}, D.~J. {Macomb}, H.~A. {Mayer-Hasselwander}, P.~F. {Michelson}, C.~{von Montigny}, R.~{Mukherjee}, P.~L. {Nolan}, P.~V. {Ramanamurthy}, E.~{Schneid}, P.~{Sreekumar}, D.~J. {Thompson}, T.~D. {Willis}, {The Likelihood Analysis of EGRET Data}, ApJ 461 (1996) 396.
\newblock \href {https://doi.org/10.1086/177068} {\path{doi:10.1086/177068}}.

\bibitem{Abdollahi_2020}
S.~Abdollahi, F.~Acero, M.~Ackermann, M.~Ajello, W.~B. Atwood, M.~Axelsson, L.~Baldini, J.~Ballet, G.~Barbiellini, D.~Bastieri, J.~B. Gonzalez, R.~Bellazzini, A.~Berretta, E.~Bissaldi, R.~D. Blandford, E.~D. Bloom, R.~Bonino, E.~Bottacini, T.~J. Brandt, J.~Bregeon, P.~Bruel, R.~Buehler, T.~H. Burnett, S.~Buson, R.~A. Cameron, R.~Caputo, P.~A. Caraveo, J.~M. Casandjian, D.~Castro, E.~Cavazzuti, E.~Charles, S.~Chaty, S.~Chen, C.~C. Cheung, G.~Chiaro, S.~Ciprini, J.~Cohen-Tanugi, L.~R. Cominsky, J.~Coronado-Blázquez, D.~Costantin, A.~Cuoco, S.~Cutini, F.~D’Ammando, M.~DeKlotz, P.~de~la Torre~Luque, F.~de~Palma, A.~Desai, S.~W. Digel, N.~D. Lalla, M.~D. Mauro, L.~D. Venere, A.~Domínguez, D.~Dumora, F.~F. Dirirsa, S.~J. Fegan, E.~C. Ferrara, A.~Franckowiak, Y.~Fukazawa, S.~Funk, P.~Fusco, F.~Gargano, D.~Gasparrini, N.~Giglietto, P.~Giommi, F.~Giordano, M.~Giroletti, T.~Glanzman, D.~Green, I.~A. Grenier, S.~Griffin, M.-H. Grondin, J.~E. Grove, S.~Guiriec, A.~K. Harding, K.~Hayashi, E.~Hays, J.~W. Hewitt,
  D.~Horan, G.~Jóhannesson, T.~J. Johnson, T.~Kamae, M.~Kerr, D.~Kocevski, M.~Kovac’evic’, M.~Kuss, D.~Landriu, S.~Larsson, L.~Latronico, M.~Lemoine-Goumard, J.~Li, I.~Liodakis, F.~Longo, F.~Loparco, B.~Lott, M.~N. Lovellette, P.~Lubrano, G.~M. Madejski, S.~Maldera, D.~Malyshev, A.~Manfreda, E.~J. Marchesini, L.~Marcotulli, G.~Martí-Devesa, P.~Martin, F.~Massaro, M.~N. Mazziotta, J.~E. McEnery, I.~Mereu, M.~Meyer, P.~F. Michelson, N.~Mirabal, T.~Mizuno, M.~E. Monzani, A.~Morselli, I.~V. Moskalenko, M.~Negro, E.~Nuss, R.~Ojha, N.~Omodei, M.~Orienti, E.~Orlando, J.~F. Ormes, M.~Palatiello, V.~S. Paliya, D.~Paneque, Z.~Pei, H.~Peña-Herazo, J.~S. Perkins, M.~Persic, M.~Pesce-Rollins, V.~Petrosian, L.~Petrov, F.~Piron, H.~Poon, T.~A. Porter, G.~Principe, S.~Rainò, R.~Rando, M.~Razzano, S.~Razzaque, A.~Reimer, O.~Reimer, Q.~Remy, T.~Reposeur, R.~W. Romani, P.~M.~S. Parkinson, F.~K. Schinzel, D.~Serini, C.~Sgrò, E.~J. Siskind, D.~A. Smith, G.~Spandre, P.~Spinelli, A.~W. Strong, D.~J. Suson, H.~Tajima, M.~N.
  Takahashi, D.~Tak, J.~B. Thayer, D.~J. Thompson, L.~Tibaldo, D.~F. Torres, E.~Torresi, J.~Valverde, B.~V. Klaveren, P.~van Zyl, K.~Wood, M.~Yassine, G.~Zaharijas, \href{https://dx.doi.org/10.3847/1538-4365/ab6bcb}{Fermi large area telescope fourth source catalog}, ApJS 247~(1) (2020) 33.
\newblock \href {https://doi.org/10.3847/1538-4365/ab6bcb} {\path{doi:10.3847/1538-4365/ab6bcb}}.
\newline\urlprefix\url{https://dx.doi.org/10.3847/1538-4365/ab6bcb}

\bibitem{2019SCPMA..6259502J}
P.~{Jiang}, Y.~{Yue}, H.~{Gan}, R.~{Yao}, H.~{Li}, G.~{Pan}, J.~{Sun}, D.~{Yu}, H.~{Liu}, N.~{Tang}, L.~{Qian}, J.~{Lu}, J.~{Yan}, B.~{Peng}, S.~{Zhang}, Q.~{Wang}, Q.~{Li}, D.~{Li}, {FAST Collaboration}, {Commissioning progress of the FAST}, Science China Physics, Mechanics, and Astronomy 62~(5) (2019) 959502.
\newblock \href {http://arxiv.org/abs/1903.06324} {\path{arXiv:1903.06324}}, \href {https://doi.org/10.1007/s11433-018-9376-1} {\path{doi:10.1007/s11433-018-9376-1}}.

\bibitem{2020RAA....20...64J}
P.~{Jiang}, N.-Y. {Tang}, L.-G. {Hou}, M.-T. {Liu}, M.~{Kr{\v{c}}o}, L.~{Qian}, J.-H. {Sun}, T.-C. {Ching}, B.~{Liu}, Y.~{Duan}, Y.-L. {Yue}, H.-Q. {Gan}, R.~{Yao}, H.~{Li}, G.-F. {Pan}, D.-J. {Yu}, H.-F. {Liu}, D.~{Li}, B.~{Peng}, J.~{Yan}, {FAST Collaboration}, {The fundamental performance of FAST with 19-beam receiver at L band}, Research in Astronomy and Astrophysics 20~(5) (2020) 064.
\newblock \href {http://arxiv.org/abs/2002.01786} {\path{arXiv:2002.01786}}, \href {https://doi.org/10.1088/1674-4527/20/5/64} {\path{doi:10.1088/1674-4527/20/5/64}}.

\bibitem{2020Innov...100053Q}
L.~{Qian}, R.~{Yao}, J.~{Sun}, J.~{Xu}, Z.~{Pan}, P.~{Jiang}, {FAST: Its Scientific Achievements and Prospects}, The Innovation 1~(3) (2020) 100053.
\newblock \href {http://arxiv.org/abs/2011.13542} {\path{arXiv:2011.13542}}, \href {https://doi.org/10.1016/j.xinn.2020.100053} {\path{doi:10.1016/j.xinn.2020.100053}}.

\bibitem{FFA_2020MNRAS.497.4654M}
V.~{Morello}, E.~D. {Barr}, B.~W. {Stappers}, E.~F. {Keane}, A.~G. {Lyne}, {Optimal periodicity searching: revisiting the fast folding algorithm for large-scale pulsar surveys}, MNRAS 497~(4) (2020) 4654--4671.
\newblock \href {http://arxiv.org/abs/2004.03701} {\path{arXiv:2004.03701}}, \href {https://doi.org/10.1093/mnras/staa2291} {\path{doi:10.1093/mnras/staa2291}}.

\bibitem{2024SCPMA..6769512Z}
D.~{Zhou}, P.~{Wang}, D.~{Li}, J.~{Fang}, C.~{Miao}, P.~C.~C. {Freire}, L.~{Zhang}, D.~{Zhang}, H.~{Chen}, Y.~{Feng}, Y.~{Xiao}, J.~{Xie}, X.~{Zhang}, C.~{Jin}, H.~{Wang}, Y.~{Ke}, X.~{Guo}, R.~{Zhao}, C.~{Niu}, W.~{Zhu}, M.~{Xue}, Y.~{Wang}, J.~{Wu}, Z.~{Gan}, Z.~{Sun}, C.~{Wang}, J.~{Zhang}, J.~{Zhang}, J.~{Cao}, W.~{Lu}, {A discovery of two slow pulsars with FAST: ``Ronin'' from the globular cluster M15}, Science China Physics, Mechanics, and Astronomy 67~(6) (2024) 269512.
\newblock \href {http://arxiv.org/abs/2312.05868} {\path{arXiv:2312.05868}}, \href {https://doi.org/10.1007/s11433-023-2362-x} {\path{doi:10.1007/s11433-023-2362-x}}.

\bibitem{2012ApJ...749...54H}
D.~{Hadasch}, D.~F. {Torres}, T.~{Tanaka}, R.~H.~D. {Corbet}, A.~B. {Hill}, R.~{Dubois}, G.~{Dubus}, T.~{Glanzman}, S.~{Corbel}, J.~{Li}, Y.~P. {Chen}, S.~{Zhang}, G.~A. {Caliandro}, M.~{Kerr}, J.~L. {Richards}, W.~{Max-Moerbeck}, A.~{Readhead}, G.~{Pooley}, {Long-term Monitoring of the High-energy {\ensuremath{\gamma}}-Ray Emission from LS I +61{\textdegree}303 and LS 5039}, ApJ 749~(1) (2012) 54.
\newblock \href {http://arxiv.org/abs/1202.1866} {\path{arXiv:1202.1866}}, \href {https://doi.org/10.1088/0004-637X/749/1/54} {\path{doi:10.1088/0004-637X/749/1/54}}.

\bibitem{2013MNRAS.436..740C}
G.~A. {Caliandro}, A.~B. {Hill}, D.~F. {Torres}, D.~{Hadasch}, P.~{Ray}, A.~{Abdo}, J.~W.~T. {Hessels}, A.~{Ridolfi}, A.~{Possenti}, M.~{Burgay}, N.~{Rea}, P.~H.~T. {Tam}, R.~{Dubois}, G.~{Dubus}, T.~{Glanzman}, T.~{Jogler}, {The missing GeV {\ensuremath{\gamma}}-ray binary: searching for HESS J0632+057 with Fermi-LAT}, MNRAS 436~(1) (2013) 740--749.
\newblock \href {http://arxiv.org/abs/1308.5234} {\path{arXiv:1308.5234}}, \href {https://doi.org/10.1093/mnras/stt1615} {\path{doi:10.1093/mnras/stt1615}}.

\bibitem{2015ApJ...811...68C}
G.~A. {Caliandro}, C.~C. {Cheung}, J.~{Li}, J.~D. {Scargle}, D.~F. {Torres}, K.~S. {Wood}, M.~{Chernyakova}, {Gamma-Ray Flare Activity from PSR B1259-63 during 2014 Periastron Passage and Comparison to Its 2010 Passage}, ApJ 811~(1) (2015) 68.
\newblock \href {http://arxiv.org/abs/1509.02856} {\path{arXiv:1509.02856}}, \href {https://doi.org/10.1088/0004-637X/811/1/68} {\path{doi:10.1088/0004-637X/811/1/68}}.

\bibitem{2012ApJ...754L..10A}
J.~{Aleksi{\'c}}, E.~A. {Alvarez}, L.~A. {Antonelli}, P.~{Antoranz}, M.~{Asensio}, M.~{Backes}, U.~{Barres de Almeida}, J.~A. {Barrio}, D.~{Bastieri}, J.~{Becerra Gonz{\'a}lez}, W.~{Bednarek}, K.~{Berger}, E.~{Bernardini}, A.~{Biland}, O.~{Blanch}, R.~K. {Bock}, A.~{Boller}, G.~{Bonnoli}, D.~{Borla Tridon}, V.~{Bosch-Ramon}, T.~{Bretz}, A.~{Ca{\~n}ellas}, E.~{Carmona}, A.~{Carosi}, P.~{Colin}, E.~{Colombo}, J.~L. {Contreras}, J.~{Cortina}, L.~{Cossio}, S.~{Covino}, P.~{Da Vela}, F.~{Dazzi}, A.~{De Angelis}, G.~{De Caneva}, E.~{De Cea del Pozo}, B.~{De Lotto}, C.~{Delgado Mendez}, A.~{Diago Ortega}, M.~{Doert}, A.~{Dom{\'\i}nguez}, D.~{Dominis Prester}, D.~{Dorner}, M.~{Doro}, D.~{Eisenacher}, D.~{Elsaesser}, D.~{Ferenc}, M.~V. {Fonseca}, L.~{Font}, C.~{Fruck}, R.~J. {Garc{\'\i}a L{\'o}pez}, M.~{Garczarczyk}, D.~{Garrido Terrats}, G.~{Giavitto}, N.~{Godinovi{\'c}}, A.~{Gonz{\'a}lez Mu{\~n}oz}, S.~R. {Gozzini}, D.~{Hadasch}, D.~{H{\"a}fner}, A.~{Herrero}, D.~{Hildebrand}, J.~{Hose}, D.~{Hrupec}, B.~{Huber},
  F.~{Jankowski}, T.~{Jogler}, V.~{Kadenius}, H.~{Kellermann}, S.~{Klepser}, T.~{Kr{\"a}henb{\"u}hl}, J.~{Krause}, A.~{La Barbera}, D.~{Lelas}, E.~{Leonardo}, N.~{Lewandowska}, E.~{Lindfors}, S.~{Lombardi}, M.~{L{\'o}pez}, R.~{L{\'o}pez-Coto}, A.~{L{\'o}pez-Oramas}, E.~{Lorenz}, M.~{Makariev}, G.~{Maneva}, N.~{Mankuzhiyil}, K.~{Mannheim}, L.~{Maraschi}, M.~{Mariotti}, M.~{Mart{\'\i}nez}, D.~{Mazin}, M.~{Meucci}, J.~M. {Miranda}, R.~{Mirzoyan}, J.~{Mold{\'o}n}, A.~{Moralejo}, P.~{Munar-Adrover}, A.~{Niedzwiecki}, D.~{Nieto}, K.~{Nilsson}, N.~{Nowak}, R.~{Orito}, S.~{Paiano}, D.~{Paneque}, R.~{Paoletti}, S.~{Pardo}, J.~M. {Paredes}, S.~{Partini}, M.~A. {Perez-Torres}, M.~{Persic}, M.~{Pilia}, J.~{Pochon}, F.~{Prada}, P.~G. {Prada Moroni}, E.~{Prandini}, I.~{Puerto Gimenez}, I.~{Puljak}, I.~{Reichardt}, R.~{Reinthal}, W.~{Rhode}, M.~{Rib{\'o}}, J.~{Rico}, S.~{R{\"u}gamer}, A.~{Saggion}, K.~{Saito}, T.~Y. {Saito}, M.~{Salvati}, K.~{Satalecka}, V.~{Scalzotto}, V.~{Scapin}, C.~{Schultz}, T.~{Schweizer}, S.~N.
  {Shore}, A.~{Sillanp{\"a}{\"a}}, J.~{Sitarek}, I.~{Snidaric}, D.~{Sobczynska}, F.~{Spanier}, S.~{Spiro}, V.~{Stamatescu}, A.~{Stamerra}, B.~{Steinke}, J.~{Storz}, N.~{Strah}, S.~{Sun}, T.~{Suri{\'c}}, L.~{Takalo}, H.~{Takami}, F.~{Tavecchio}, P.~{Temnikov}, T.~{Terzi{\'c}}, D.~{Tescaro}, M.~{Teshima}, O.~{Tibolla}, D.~F. {Torres}, A.~{Treves}, M.~{Uellenbeck}, P.~{Vogler}, R.~M. {Wagner}, Q.~{Weitzel}, V.~{Zabalza}, F.~{Zandanel}, R.~{Zanin}, {Detection of VHE {\ensuremath{\gamma}}-Rays from HESS J0632+057 during the 2011 February X-Ray Outburst with the MAGIC Telescopes}, ApJL 754~(1) (2012) L10.
\newblock \href {http://arxiv.org/abs/1203.2867} {\path{arXiv:1203.2867}}, \href {https://doi.org/10.1088/2041-8205/754/1/L10} {\path{doi:10.1088/2041-8205/754/1/L10}}.

\bibitem{1992ApJ...387L..37J}
S.~{Johnston}, R.~N. {Manchester}, A.~G. {Lyne}, M.~{Bailes}, V.~M. {Kaspi}, G.~{Qiao}, N.~{D'Amico}, {PSR 1259-63: A Binary Radio Pulsar with a Be Star Companion}, ApJL 387 (1992) L37.
\newblock \href {https://doi.org/10.1086/186300} {\path{doi:10.1086/186300}}.

\bibitem{2009ApJ...705....1C}
F.~{Camilo}, P.~S. {Ray}, S.~M. {Ransom}, M.~{Burgay}, T.~J. {Johnson}, M.~{Kerr}, E.~V. {Gotthelf}, J.~P. {Halpern}, J.~{Reynolds}, R.~W. {Romani}, P.~{Demorest}, S.~{Johnston}, W.~{van Straten}, P.~M. {Saz Parkinson}, M.~{Ziegler}, M.~{Dormody}, D.~J. {Thompson}, D.~A. {Smith}, A.~K. {Harding}, A.~A. {Abdo}, F.~{Crawford}, P.~C.~C. {Freire}, M.~{Keith}, M.~{Kramer}, M.~S.~E. {Roberts}, P.~{Weltevrede}, K.~S. {Wood}, {Radio Detection of LAT PSRs J1741-2054 and J2032+4127: No Longer Just Gamma-ray Pulsars}, ApJ 705~(1) (2009) 1--13.
\newblock \href {http://arxiv.org/abs/0908.2626} {\path{arXiv:0908.2626}}, \href {https://doi.org/10.1088/0004-637X/705/1/1} {\path{doi:10.1088/0004-637X/705/1/1}}.

\bibitem{2022NatAs...6..698W}
S.-S. {Weng}, L.~{Qian}, B.-J. {Wang}, D.~F. {Torres}, A.~{Papitto}, P.~{Jiang}, R.~{Xu}, J.~{Li}, J.-Z. {Yan}, Q.-Z. {Liu}, M.-Y. {Ge}, Q.-R. {Yuan}, {Radio pulsations from a neutron star within the gamma-ray binary LS I +61{\textdegree} 303}, Nature Astronomy 6 (2022) 698--702.
\newblock \href {http://arxiv.org/abs/2203.09423} {\path{arXiv:2203.09423}}, \href {https://doi.org/10.1038/s41550-022-01630-1} {\path{doi:10.1038/s41550-022-01630-1}}.

\bibitem{Emmering1989ApJ...345..931E}
R.~T. {Emmering}, R.~A. {Chevalier}, {The Intrinsic Luminosity and Initial Period of Pulsars}, ApJ 345 (1989) 931.
\newblock \href {https://doi.org/10.1086/167963} {\path{doi:10.1086/167963}}.

\bibitem{Kijak2003A&A...397..969K}
J.~{Kijak}, J.~{Gil}, {Radio emission altitude in pulsars}, A\&A 397 (2003) 969--972.
\newblock \href {https://doi.org/10.1051/0004-6361:20021583} {\path{doi:10.1051/0004-6361:20021583}}.

\bibitem{Takata2011ApJ...726...44T}
J.~{Takata}, Y.~{Wang}, K.~S. {Cheng}, {Population Study for {\ensuremath{\gamma}}-ray Pulsars with the Outer Gap Model}, ApJ 726~(1) (2011) 44.
\newblock \href {http://arxiv.org/abs/1010.5870} {\path{arXiv:1010.5870}}, \href {https://doi.org/10.1088/0004-637X/726/1/44} {\path{doi:10.1088/0004-637X/726/1/44}}.

\bibitem{chen2021A&A...652A..39C}
A.~M. {Chen}, Y.~D. {Guo}, Y.~W. {Yu}, J.~{Takata}, {Radio absorption in high-mass gamma-ray binaries}, A\&A 652 (2021) A39.
\newblock \href {http://arxiv.org/abs/2106.10445} {\path{arXiv:2106.10445}}, \href {https://doi.org/10.1051/0004-6361/202140951} {\path{doi:10.1051/0004-6361/202140951}}.

\bibitem{waters1988A&A...198..200W}
L.~B.~F.~M. {Waters}, E.~P.~J. {van den Heuvel}, A.~R. {Taylor}, G.~M.~H.~J. {Habets}, P.~{Persi}, {Evidence for low-velocity winds in Be/X-ray binaries.}, A\&A 198 (1988) 200--210.

\bibitem{wind_2020ApJ...888..115A}
A.~{Archer}, W.~{Benbow}, R.~{Bird}, A.~{Brill}, R.~{Brose}, M.~{Buchovecky}, J.~L. {Christiansen}, A.~J. {Chromey}, W.~{Cui}, A.~{Falcone}, Q.~{Feng}, J.~P. {Finley}, L.~{Fortson}, A.~{Furniss}, A.~{Gent}, G.~H. {Gillanders}, C.~{Giuri}, O.~{Gueta}, D.~{Hanna}, T.~{Hassan}, O.~{Hervet}, J.~{Holder}, G.~{Hughes}, T.~B. {Humensky}, P.~{Kaaret}, N.~{Kelley-Hoskins}, M.~{Kertzman}, D.~{Kieda}, M.~{Krause}, M.~J. {Lang}, G.~{Maier}, P.~{Moriarty}, R.~{Mukherjee}, D.~{Nieto}, M.~{Nievas-Rosillo}, S.~{O'Brien}, R.~A. {Ong}, A.~N. {Otte}, N.~{Park}, A.~{Petrashyk}, K.~{Pfrang}, M.~{Pohl}, R.~R. {Prado}, E.~{Pueschel}, J.~{Quinn}, K.~{Ragan}, P.~T. {Reynolds}, D.~{Ribeiro}, G.~T. {Richards}, E.~{Roache}, I.~{Sadeh}, M.~{Santander}, S.~{Schlenstedt}, G.~H. {Sembroski}, I.~{Sushch}, A.~{Weinstein}, P.~{Wilcox}, A.~{Wilhelm}, D.~A. {Williams}, T.~J. {Williamson}, C.~J. {Hailey}, S.~{Mandel}, K.~{Mori}, {Probing the Properties of the Pulsar Wind in the Gamma-Ray Binary HESS J0632+057 with NuSTAR and VERITAS
  Observations}, ApJ 888~(2) (2020) 115.
\newblock \href {http://arxiv.org/abs/1911.09434} {\path{arXiv:1911.09434}}, \href {https://doi.org/10.3847/1538-4357/ab59de} {\path{doi:10.3847/1538-4357/ab59de}}.

\bibitem{2016A&A...593A..97Z}
R.~K. {Zamanov}, K.~A. {Stoyanov}, J.~{Mart{\'\i}}, G.~Y. {Latev}, Y.~M. {Nikolov}, M.~F. {Bode}, P.~L. {Luque-Escamilla}, {Optical spectroscopy of Be/gamma-ray binaries}, A\&A 593 (2016) A97.
\newblock \href {http://arxiv.org/abs/1605.05811} {\path{arXiv:1605.05811}}, \href {https://doi.org/10.1051/0004-6361/201628735} {\path{doi:10.1051/0004-6361/201628735}}.

\bibitem{extended_radio_2011A&A...533L...7M}
J.~{Mold{\'o}n}, M.~{Rib{\'o}}, J.~M. {Paredes}, {Revealing the extended radio emission from the gamma-ray binary HESS J0632+057}, A\&A 533 (2011) L7.
\newblock \href {http://arxiv.org/abs/1108.0437} {\path{arXiv:1108.0437}}, \href {https://doi.org/10.1051/0004-6361/201117764} {\path{doi:10.1051/0004-6361/201117764}}.

\end{thebibliography}






\end{document}